\documentclass[preprint,12pt]{elsarticle}
\usepackage{cite,natbib}
\usepackage{subcaption}
\usepackage{amsmath}
\usepackage{deluxetable}

\newcommand{\fun}[1]{(Mg$_2$SiO$_4$)$_{#1}$}
\newcommand{\fufig}[2]{\includegraphics[width=6em]{figures/fus/#1} n = #2}
\newcommand{\angstrom}{\text{\normalfont\AA}\space}

\begin{document}
\begin{frontmatter}
	
\title{The formation of astrophysical Mg-rich silicate dust}
\author[labmauney]{Christopher M. Mauney\fnref{osuref}}
\author[lablazzati]{Davide Lazzati\fnref{osuref}}
\address[labmauney]{mauneyc@oregonstate.edu}
\address[lablazzati]{lazzatid@science.oregonstate.edu}
\fntext[osuref]{Oregon State University}

\begin{abstract}
We present new results for ground-state candidate energies of Mg-rich olivine (MRO) clusters and use the binding energies of these clusters to determine their nucleation rates in stellar outflows, with particular interest in the environments of core-collapse supernovae (CCSNe). Low-lying structures of clusters \fun{n} $2 \le n \le 13$ are determined from a modified minima hopping algorithm using an empirical silicate potential in the Buckingham form. These configurations are further refined and optimized using the density functional theory code Quantum Espresso. Utilizing atomistic nucleation theory, we determine the critical size and nucleation rates of these clusters. We find that configurations and binding energies in this regime are very dissimilar from those of the bulk lattice. Clusters grow with SiO$_4$-MgO layering and exhibit only global, rather than local, symmetries. When compared to classical nucleation theory we find suppressed nucleation rates at most temperatures and pressures, with enhanced nucleation rates at very large pressures. This implies a slower progression of silicate dust formation in stellar environments than previously assumed.
\end{abstract}

\end{frontmatter}

\section{Introduction}
Cosmic dust presents an intriguing laboratory to the physicist. The cycle of dust, it's crucial role in a great galactic recycling and processing of material, necessitates a broad view. Yet any approach to the study of dust that does not incorporate the details of small-scale chemistry and kinetics, up to and including quantum effects, will fail as a predictive model. This is especially true when examining dust formation, when the physics is in the most dynamic phase of the cycle.

Silicate dust is a major component of dust present in the ISM. Absorption features at 9.7 $\mu$ m and 18 $\mu$ m are associated with the Si-O stretching and O-Si-O bending modes in silicates \citep{gibb2004interstellar}. Similar spectral features have been observed in other galaxies \citep{teplitz2006silicate,roche2007silicate}. These lines are strong and broad, indicating that in the diffuse ISM silicates are structurally amorphous \citep{kemper2004absence}. However, crystalline silicate features have been observed around AGB stars and stars with disks \citep{molster2002crystalline,olofsson2009c2d}, and in comets \citep{wooden1999silicate}. These observations indicate that a substantial fraction of silicate dust grains are in a crystal structure before being injected into the ISM. It is possible that crystallization occurs directly from the vapor, and subsequent processing by grain-grain collisions, shock sputtering, and thermal annealing leads to amorphization \citep{henning2010cosmic}. However, the reverse is also possible, where amorphous silicate is formed from the vapor and later processed into crystalline structure.

%Theoretical models are in disagreement on dust formation history. Early studies based on classical nucleation theory (CNT) found dust mass yields of $\approx 1$ \msol \citep{Kozasa89,Kozasa91,Nozawa03,Nozawa07,Fallest11,Todini01}. Models based in CNT will produce nearly the entirety of expected dust mass in a "flash" about 100 days after explosion. An alternate approach, based on chemical kinetics, better tracks dust formation history but tends to dust yields that are magnitudes below CNT predictions \citep{Cherchneff09,Cherchneff10}. To enhance the dust production of kinetics, Sarangi \& Cherchneff \citep{Sarangi15} implemented a scheme for coagulation of large molecules coupled with gas-phase chemistry and found it to be an efficient mechanism of dust growth. However, this method is dependent on the formation of relatively large dust precursors in the gas phase chemistry, and further assumes that the critical cluster size is the largest molecule in the network independent of temperature and gas density. 

Determining the formation pathway of silicate dust grains is necessary for making more accurate predictions of dust properties. In this paper we continue the approach set forth in our previous paper on carbon dust precursors \citep{Mauney15} with the Mg-rich olivine (MRO) clusters \fun{n}, for $2 < n < 13$. In principle, to exactly determine nucleation of multi-species molecules, all pathways through a high-dimensional Gibbs energy surface are necessary. For the reasons given in the next section, we instead follow a fixed stochiometric ratio into larger molecules.

This paper is organized as follows: in section \ref{sec:methods} we detail our methods for cluster ground state configuration and density functional theory energy calculations. Results of these calculations are presented in section \ref{sec:results}, showing binding energies of clusters, critical sizes, and nucleation rates. We summarize these results and discuss their impact in section \ref{sec:conclusion}.

\section{Methods}
\label{sec:methods}
\subsection{Selecting the nucleation pathway}

In principle the formation pathway of silicates can take many directions through precursor molecules that do not necessarily maintain a fixed stoichiometry. It is to be expected, however, that the stoichiometry of the precursor molecule will eventually approach the one of the crystal. Since it is computationally impractical to compute the properties of all possible precursor molecules, in this study we only investigate molecules of the form \fun{n}. While this is a limitation of this work, it is worth noting that it is still a significant improvement over the capillary approximation used in other investigations of silicate nucleation (see, e.g., \citep{Todini01,Nozawa03}).  Not only the fixed stoichiometry is implicitly assumed, but also the precursor molecules are taken to be spherical and possessing the physical properties - such as surface energy - of the bulk material. While some studies \citep{Goumans12,Bromley16} have attempted a full study of precursor molecules of free stoichiometry, they were limited to small molecules ($n < 10$) (the larger molecules in those studies did follow fixed stoichiometries).

\subsection{Determining ground state configuration}
A molecular configuration approximating the ground state of the molecule must be supplied to the DFT calculation. We use an implementation of the global minima search algorithm minima-hopping (MH) developed by Goedecker \citep{Goedecker04}. Minima hopping does not generate new configurations based on
random moves, as other global minima techniques such as basin hopping, but rather smoothly follows the energy surface by
applying molecular dynamics to the system.  Starting from an initial
state, the system is given a kinetic energy and allowed to evolve
according to the equations of motion. The stopping criterion of the
molecular dynamics algorithm is passing over one or two hills on the potential energy surface (that is, the system goes over
from increasing to deceasing energy).

New configurations are subjected to a
minimization after the molecular dynamics step. The minimized configuration is compared to the previous (beginning) one.
If the two configurations are determined to be the same (in this case, comparing inter-atomic distances of atoms), the algorithm returns to performing the molecular dynamics
step with the an increased kinetic energy. If the system escapes the current energy well
into a new unique one (that is, a new configuration is found), the kinetic energy is reduced, and the process is started again. Details on the specifics of implementation of the algorithm can be found in \citep{Goedecker04}. All the parameters we used in this article can be found in Table \ref{tbl:mh_parameters_0}.

We use a modified version of minima hopping that implements an atom-swap. This algorithm defines a small (1-5\%) probability $\alpha$ that the next cycle will swap positions of two randomly selected atoms in the molecule rather than do a molecular dynamics run. This use of non-local transformations has been applied to basin-hopping and has been found to increase the efficiency of such searches \citep{rondina2013revised}. In our approach, a temperature $T_s$ is used in a simple Metropolis condition $\exp(-\Delta E / T_s)$ after the swap to determine if the swapped positions should be accepted as the new state of the search, where $\Delta E$ represents the relative change in energy.

The potential function used in this study is the Bees-Kramer-van Santen (BKS) model \citep{van1990force}. The fuctional form is given as a combination of Coulomb and Buckingham potential terms

\begin{equation}
\begin{split}
\label{eq:buckingham_potential_0}
U(\mathbf{r}) &= \sum_{i < j} U_{ij} + \sum_{i < j < k} U_{ijk}\\
U_{ij} &= \frac{q_i q_j}{r_{ij}} + A_{ij}\exp (-r_{ij}/B_{ij}) - C_{ij} r_{ij}^{-6}\\
U_{ijk}&= K_{ijk} \left(\theta_{ijk} - \theta_{0,ijk}\right)^2
\end{split}
\end{equation}
where the variable $r_{ij}$ is the inter-atomic distance between atoms $i$ and $j$ and $q_i$ is the charge on atom $i$. The pair-wise parameters $A_{ij}$ and $B_{ij}$ specify a short range repulsive force and $C_{ij}$ a long range attractive force. The three-body parameter $K_{ijk}$ is a force constant, $\theta_{ijk}$ is the angle formed from atoms $i,j,k$, and $\theta_0$ is the equilibrium angle. Values for these parameters are determined from ab-inito studies and reliably reproduce the properties of crystalline and large amorphous silicates. The values used in this study are given in Tables \ref{tbl:charge_potential_parameters_0}, \ref{tbl:pair_potential_parameters_0}, \& \ref{tbl:angular_potential_parameters_0} \citep{hassanali2007model,roberts2001investigation,flikkema2003new}.

Table \ref{tbl:mh_parameters_0} gives the values used for our minima hopping parameters. Verlet integration is used to evolve the system during the molecular dynamics step of the minima hopping algorithm. A Broyden-Fletcher-Goldfarb-Shanno (BFGS) \citep{shanno1985broyden} algorithm is used for the minimization step.

\begin{deluxetable}{ll}
	\tablecolumns{2}
	\tablewidth{0pc}
	\tablecaption{Charges of elements}
	\tablehead{
		\colhead{atom} & \colhead{charge ($e$)} } 
	\startdata
Mg & +2.0 \\
O & -2.0 \\
Si & +4.0 \\
	\enddata
	\label{tbl:charge_potential_parameters_0}
\end{deluxetable}

\begin{deluxetable}{llll}
\tablecolumns{4}
\tablewidth{0pc}
\tablecaption{Pair-potential parameters for the Buckingham potential}
\tablehead{
	\colhead{pair} & \colhead{$A_{ij} (eV)$} & \colhead{$B_{ij} (\angstrom^{-1})$} & \colhead{$C_{ij} (eV \angstrom^{-6})$}  }
\startdata
O-O & 22764.0 & 1/0.149 & 27.88 \\
Si-O & 1283.907 & 1/0.32052 & 10.66158 \\ %\hline
Si-Si & 79502.113 & 1/0.201 & 446.780 \\ %\hline
Mg-O & 821.6 & 1/0.3242 & 0.0 \\
\enddata
\label{tbl:pair_potential_parameters_0}
\end{deluxetable}

%\begin{table}
%	\centering
%	\begin{tabular}{| l | l | l | l |}
%		\hline
%		pair & A (eV) & B (\angstrom $^{-1}$) & C (eV \angstrom $^{-6}$) \\ \hline
%		O-O & 22764.0 & 1/0.149 & 27.88 \\
%		Si-O & 1283.907 & 1/0.32052 & 10.66158 \\ %\hline
%		Si-Si & 79502.113 & 1/0.201 & 446.780 \\ %\hline
%		Mg-O & 821.6 & 1/0.3242 & 0.0 \\ \hline
%		species & charge & & \\ \hline
%		Mg & 2.0 & & \\
%		O & -2.0 & & \\
%		Si & 4.0 & & \\ \hline
%	\end{tabular}
%	\caption{pair-potential parameters}
%	\label{tbl:pair_potential_parameters_0}
%\end{table}

\begin{deluxetable}{llll}
	\tablecolumns{3}
	\tablewidth{0pc}
	\tablecaption{Three-body potential parameters}
	\tablehead{
		\colhead{tuple} & \colhead{$K_{ijk} (eV \angstrom^{-2})$} & \colhead{$\theta_{0,ijk}$}  }
	\startdata
	O-Si-O & 2.097 & 109.47$^{\circ}$ \\
    O-Mg-O & 2.097 & 90.0$^{\circ}$ \\	
	\enddata
	\label{tbl:angular_potential_parameters_0}
\end{deluxetable}

\begin{deluxetable}{lllllll}
	\tablecolumns{7}
	\tablewidth{0pc}
	\tablecaption{Minima hopping parameters}
	\tablehead{
		\colhead{$\alpha_1$} & \colhead{$\alpha_2$} & \colhead{$\beta_1$} & \colhead{$\beta_2$} & \colhead{$\beta_3$} & \colhead{$E^0_{diff} (eV)$} & \colhead{$T_0 (K)$} 
	}
	\startdata
	0.95 & 1.05 & 1.05 & 1.05 & 0.95 & 0.5 & 1000
	\enddata
	\label{tbl:mh_parameters_0}
\end{deluxetable}

%\begin{table}
%	\centering
%	\begin{tabular}{| l | l | l |}
%		\hline
%		tuple & K (eV \angstrom$^{-2}$) & $\theta_0$ \\ \hline
%		O-Si-O & 2.097 & 109.47$^{\circ}$ \\
%		O-Mg-O & 2.097 & 90.0$^{\circ}$ \\ \hline		
%	\end{tabular}
%	\caption{three-body parameters}
%	\label{tbl:angular_potential_parameters_0}
%\end{table}

\begin{figure}[!ht]
	\centering
	\includegraphics[width=1.0\textwidth]{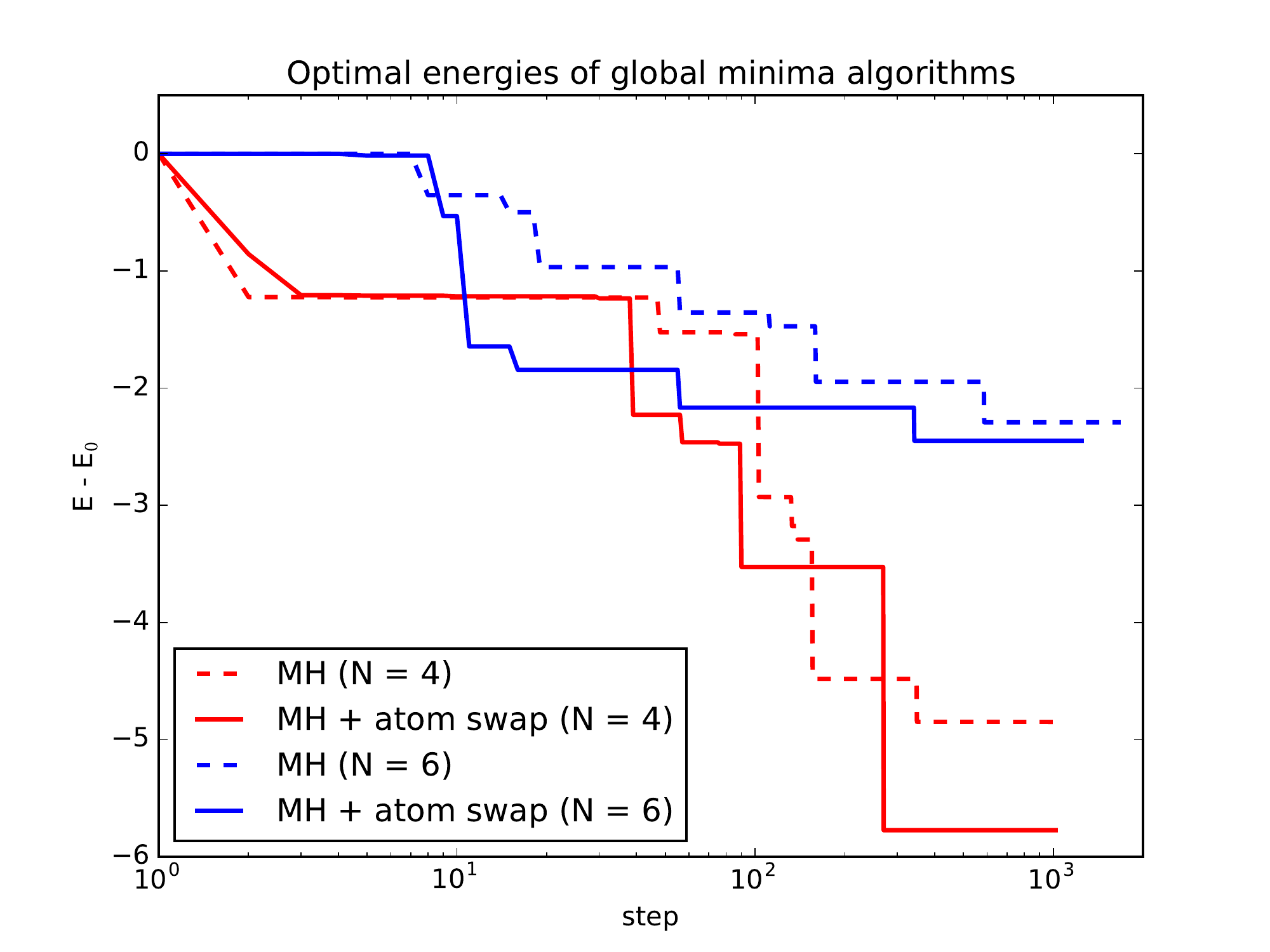}
	\caption{Performance of traditional minima-hopping algorithms (dashed) compared with minima-hopping with atom-swap included (solid), using the same initial starting configuration.}
	\label{fig:mh_as_compare}
\end{figure}

\begin{figure}[!ht]
	\centering
	\includegraphics[width=1.0\textwidth]{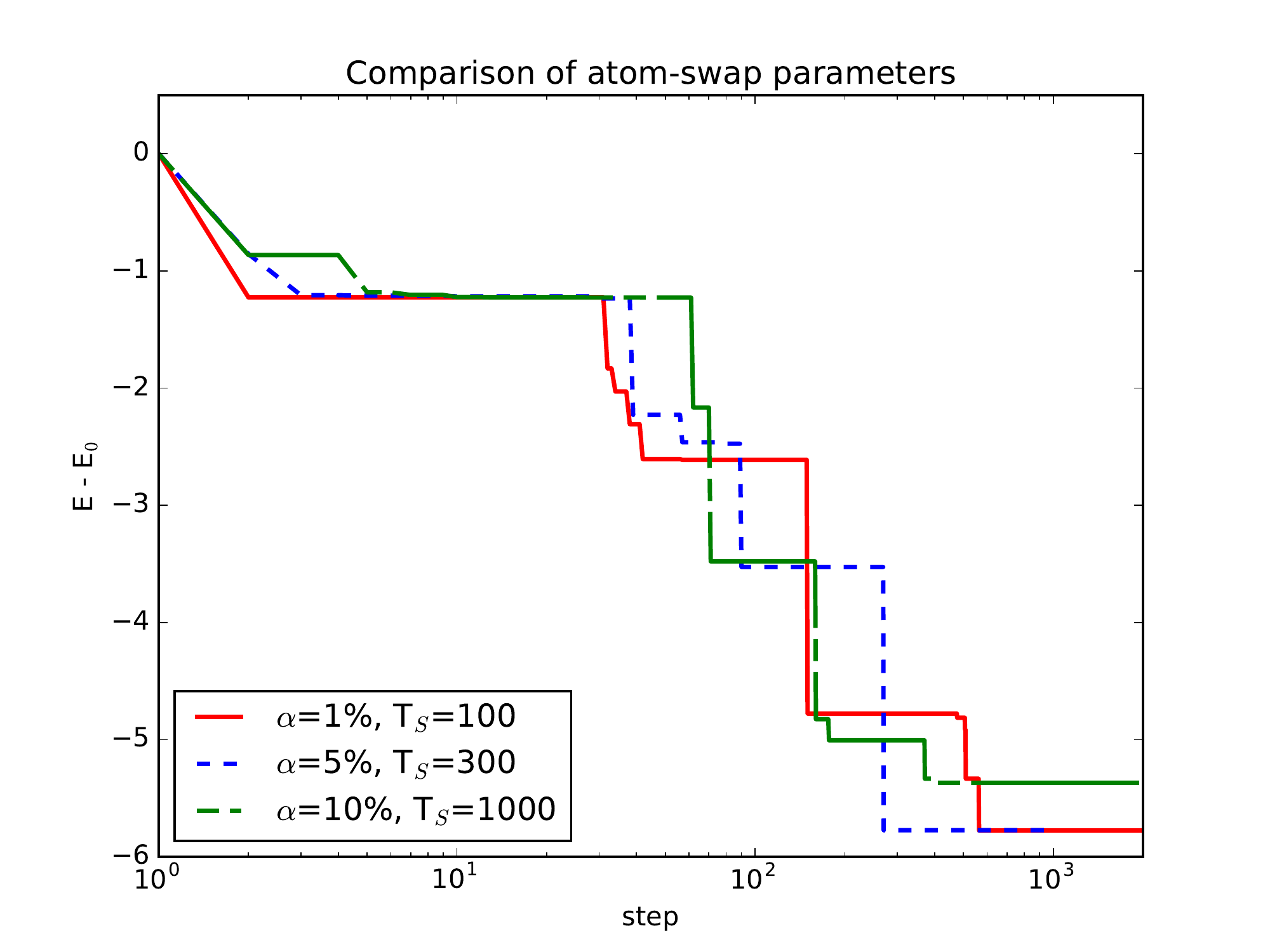}
	\caption{Comparison of different atom-swap parameters (swap \% and swap temperature) for the same initial starting configuration.}
	\label{fig:mh_as_param_compare}
\end{figure}

Our atom-swap method results in most cases in faster searches. Figure \ref{fig:mh_as_compare} shows a comparison between traditional minima hopping and minima hopping with atom-swap for a $n=4$ and $n=6$ test case. In both cases, the atom-swap method yields lower energies. Figure \ref{fig:mh_as_param_compare} compares atom-swap algorithms with different parameters. The algorithm is not very dependent on the selection of parameters, but appears to work best at $\alpha=5 \%$ and $T=300$. We select these values for this article.

\begin{figure}[!ht]
	\centering
	\includegraphics[width=1.0\textwidth]{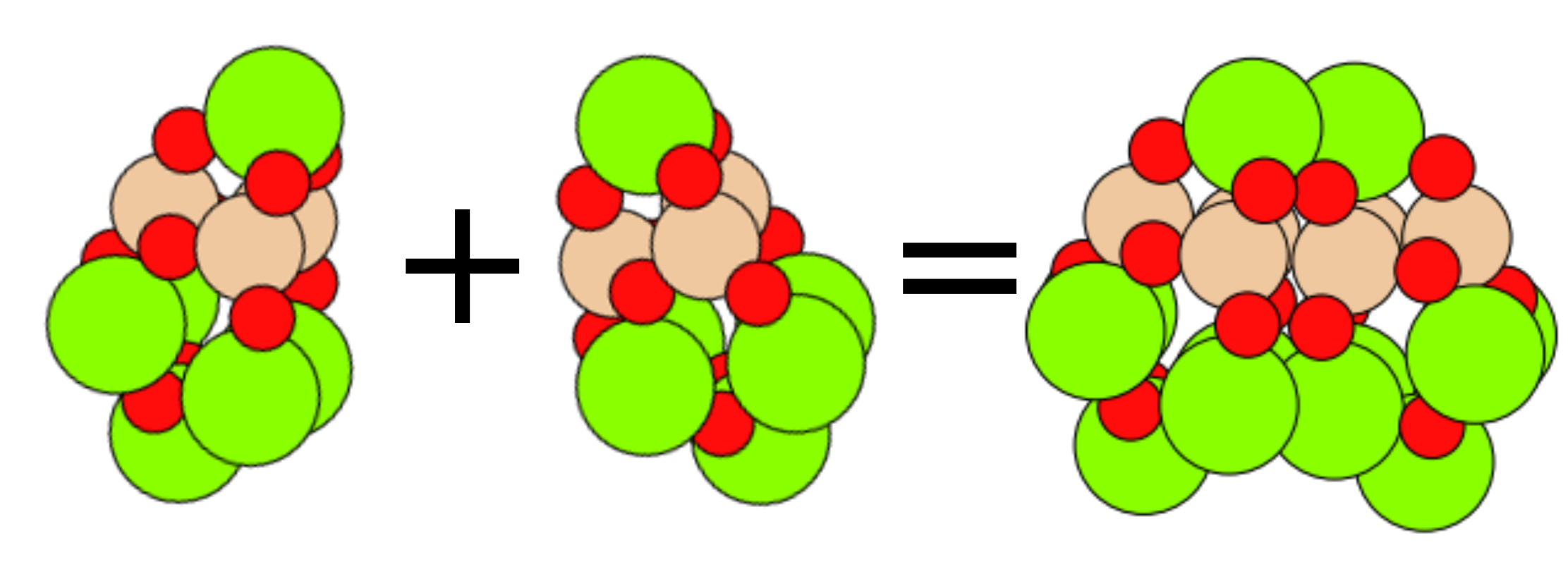}
	\caption{Example procedure of generating starting configurations of larger cluster searches.}
	\label{fig:mh_combine_0}
\end{figure}

The number of discoverable minima increases exponentially with the size of the molecule. It is therefore advantageous, especially for larger silicate clusters, to precondition the input to the minima-hopping algorithm. For larger molecules ($n > 5$), we borrow techniques from genetic algorithms used in structure prediction \citep{Deaven95}, where previously determined clusters can act as sub-units in the seeds of larger clusters. Inputs are constructed as combinations of smaller molecules found in previous runs. An example is provided in Figure \ref{fig:mh_combine_0}, where we use the optimized state of \fun{3} to generate the initial search state of the \fun{6} molecule.

%So, for example, the input configuration of \fun{6} is given as two \fun{3} molecules "stiched" together. This procedure requires some finesse, but results in a much more efficient minima search.

%To this end, we use as input a pre-optimized Lennard-Jones (LJ) cluster for which a global minima has previously been found (cite).  Each LJ atom in the input is assigned an atomic species until all necessary species for the particular molecule are used.

It is possible that the lowest-lying state produced using an empirical potential will not be the lowest in the DFT calculation. To address this, we select 8 lowest-lying candidate configurations from the global search. These are screened using a low-resolution DFT relaxation, and the candidate with the lowest reported energy is selected for a full DFT calculation.

Many of our larger \fun{n} molecules appear amorphous and lack clear symmetries. The potential given by Eq.(\ref{eq:buckingham_potential_0}) can be dominated by electrostatic forces between the Mg$^{2+}$ anions and SiO$_4^{4-}$ cations, limiting the manifestation of large-scale symmetries arising from chemical bonding. However, the thoroughness of our search and robustness of the MH algorithm gives us confidence that our configurations are energetically low-lying states.

\subsection{Quantum chemistry calculations of \fun{n} clusters}

We use the DFT software Quantum ESPRESSO v5.3 (QE) \citep{giannozzi2009quantum} to calculate the binding energies of the \fun{n} clusters. What follows is an overview of DFT and the plane-wave approach of QE uses to solve the coupled Kohn-Sham (KS) equations. For a more thorough exploration of the theory, we refer the reader to \citep{Mauney15,MartinTB}. An initial guess of the many-body electron density $n(\{\mathbf{r_i}\})$, as a function of the electron positions $\{\mathbf{r_i}\} = (\mathbf{r_1},\mathbf{r_2},...,\mathbf{r_N})$, is made and is used to construct an effective potential $V_{eff}([n])$. An updated density $n'(\{\mathbf{r_i}\})$ is found by solving the Kohn-Sham eigenvalue equations

\begin{equation}
\left[\frac{-\hbar^2}{2 m_e} \nabla_i^2 + V_{eff} \right]\phi_i = \epsilon_i \phi_i
\end{equation}
where $\hbar$ is Planks reduced constant, $m_e$ is the electron mass, $\phi_i$ are the KS orbital basis functions and $\epsilon_i$ are the eigenvalues. The new density is recovered as 

\begin{equation}
n'(\{\mathbf{r_i}\}) = \sum_i | \phi_i |^2
\end{equation}
This iterative procedure is done until self-consistency is achieved, that is when $|n - n'| \le \delta$, with $\delta$ a convergence parameter close to zero. We will be subjecting our clusters to local minimization using BFGS, which can be very computationally expensive. Therefore we set $\delta = 10^{-6}$ to allow for an efficient minimization and still maintain good accuracy of the energy. 

The binding energy $E_b$ is given as

\begin{equation}
E_b = E_t - n E_1
\label{eq:binding_energy_0}
\end{equation}
where $E_t$ is the total energy and $E_1$ is the single monomer total energy. The difference represents all energy released from the system following cluster formation.

\subsection{Free energy and nucleation}
To calculate nucleation rates, we first need to determine the change in free energy when clusters grow. We can construct the free energy of a cluster of size $n$ as

\begin{equation}
 G(n) =  G_V(n) +  G_S(n)
\label{eqn:gfe_0}
\end{equation}
where $G_V(n)$ is referred to as the volume term, and $G_S(n)$ as the surface term. These represent, respectively, the energy release when moving vapor monomers to the new phase, and the energy barrier necessary to overcome when doing so. The volume term comes from familiar thermodynamics

\begin{equation}
G_V(n) = -n k T \ln S
\label{eqn:gfe_V}
\end{equation}
with supersaturation ratio (hereafter saturation) $S = p/p_e$ where $p_e$ represents the equilibrium pressure, and temperature $T$. Following \citep{Kozasa87}, we take the pressure as the partial pressure of the key species in the formation of the clusters. The key species represents the constituent element of the vapor with the lowest collision rate, and to good approximation the rate of nucleation is controlled by the density of the key species. For MRO nucleation in astrophysical conditions we find Mg to be the limiting element, in most cases. The equilibrium pressure is given as function of temperature $T$ as

\begin{equation}
\ln (p_e) = -A/T + B
\end{equation}
where $A$,$B$ are fitted thermodynamic constants. We use the values by \citep{Nozawa03}, $A=18.62 \times 10^4 K$ and $B=52.4336$.

In the classical case of CNT, the capillary approximation is used to represent the surface term. In the atomistic case, we use

\begin{equation}
G_S(n) = \lambda n - E_n
\label{eqn:gfe_S}
\end{equation}
where $E_n$ is the binding energy of a cluster of size $n$ and $\lambda n$ represents the binding energy of $n$ monomers in the \textit{bulk} solid phase (e.g. the infinite lattice). 

For a full atomistic formulation, we go from $n$ continuous to $n$ discrete, and from Eq. (\ref{eqn:gfe_0} -\ref{eqn:gfe_S}) construct the work of cluster formation $W_n$
\begin{equation}
W_n = -n k T \ln S - (E_n - \lambda n)
\label{eqn:wcf}
\end{equation}
The maximum value of $W_n$ represents the critical cluster size; values at the critical size are denoted with a $*$, so that the critical size is $n^*$, the WCF is $W^*$, ect. To very good approximation the stationary nucleation rate $J_s$ is only a function of critical values

\begin{equation}
J_s = z f^* C^*
\label{eq:jrate}
\end{equation}
where $z$ is the Zel'dovich factor, $f^*$ is the attachment rate onto critical clusters and $C^*$ is the concentration of critical clusters. The Zel'dovich factor accounts for the possibility that critical clusters will spontaneously lose a monomer and decay into smaller clusters, rather than grow into the new phase. For the purposes of this work we take $z=1$ (that is, all critical clusters will grow into the new phase).

\subsection{Determination of $E_1$ and $\lambda$}
The properties of the molecule \fun{1}, particularly the binding energy $E_1$ and the bulk lattice cohesive energy $\lambda$ of a forsterite lattice with respect to \fun{1}, are necessary for our calculations. There exists robust literature on the chemistry of forsterite crystals, but little of it explores possible values of $E_1$ and there is disagreement on the value $\lambda$.

To determine the structure of $E_1$, we take a single \fun{1} from the bulk lattice consisting of the octahedral sites M1, M2, and the SiO$_4$ belonging to sites O2 and O3. This molecule then undergoes ionic relaxation in DFT. The ground states of the constituent atoms Mg, Si, and O are then calculated. The binding energy $E_1$ is given by

\begin{equation}
E_1 = E_{T} - 2 E_{Mg} - E_{Si} - 4 E_{O}
\label{eq:e_1}
\end{equation}
The calculation of $\lambda$ is similar. To set up the lattice, four forsterite molecules are arranged in a supercell with an orthohombic lattice, and then relaxed in DFT as above. The energy $\lambda$ is then
\begin{equation}
\lambda = (E_{lattice} - 4 E_1) / 4
\label{eq:lambda_0}
\end{equation}

Because in the bulk silicates may be either crystalline or amorphous it would be better to use a $\lambda$ reduced from the bulk crystalline to be able to explore amorphous growth. Studies of amorphous and crystalline material properties demonstrate that amorphous material has a lower bulk binding energy than the crystalline form (e.g. \citep{thogersen2008experimental,gonccalves2016molecular}). Given the configurations of the MRO clusters in Figure \ref{fig:fu_table}, it is reasonable to conclude that MRO grains grow amorphously. Therefore we select value that is lower from our crystalline lattice calculation but still within the range of other studies. In the appendix we explore how different values of $\lambda$ impact our results.

\begin{table}
	\centering
	\begin{tabular}{| c | c |}
		\hline
		$E_1$ & -389.6 Ry \\ \hline
		$\lambda$ (QE) & 9.5 \\ \hline
		$\lambda$ (ref) & \shortstack{7.4 \citep{Catti81} \\ 9.4 \citep{May00}} \\ \hline
		$\lambda$ (this work) & 8.0 \\ \hline
	\end{tabular}
	\caption{$E_1$ and $\lambda$.}
	\label{tbl:e1_lambda_values}
\end{table}

\section{Results}
\label{sec:results}
\subsection{Ground state configurations of (Mg$_2$SiO$_4$)$_n$ clusters}
\begin{figure}
	\centering
	\begin{tabular}{ccc}
		\fufig{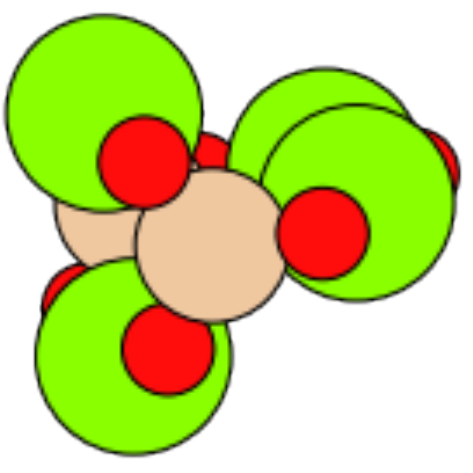}{2} & \fufig{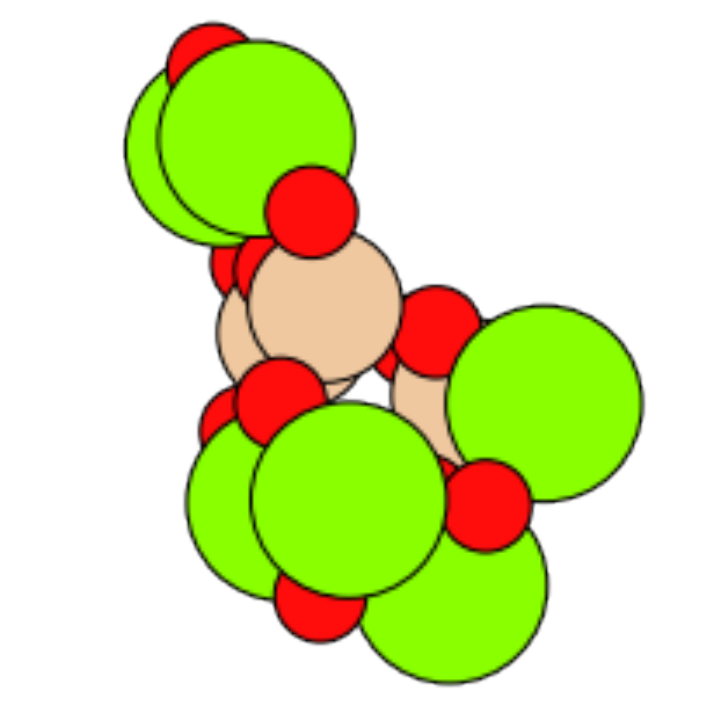}{3} & \fufig{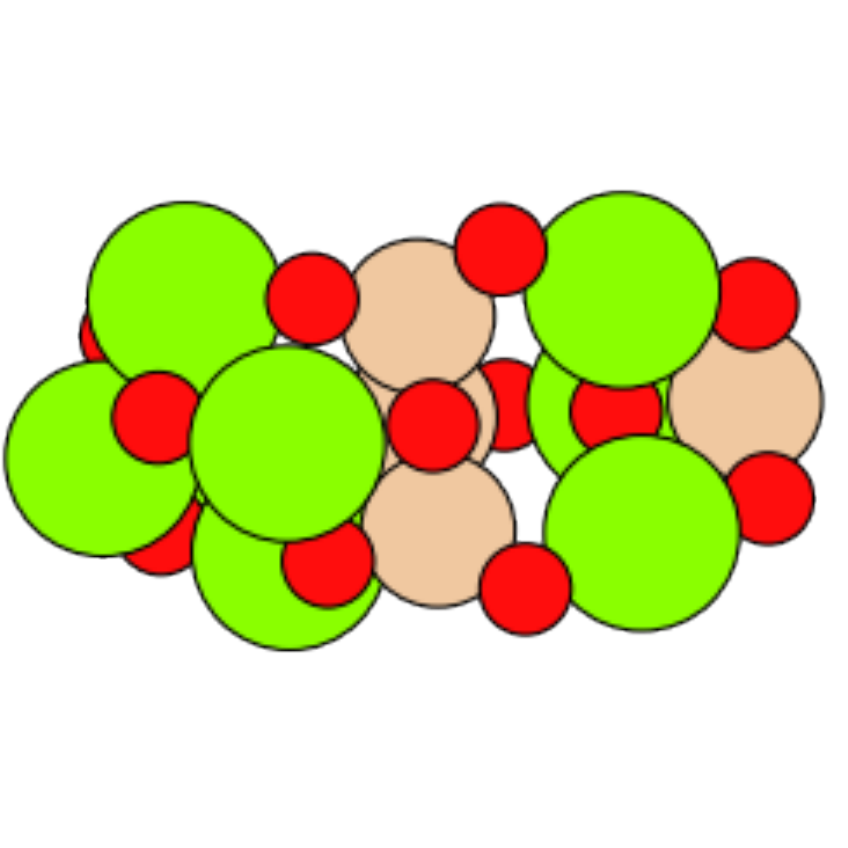}{4} \\
		\fufig{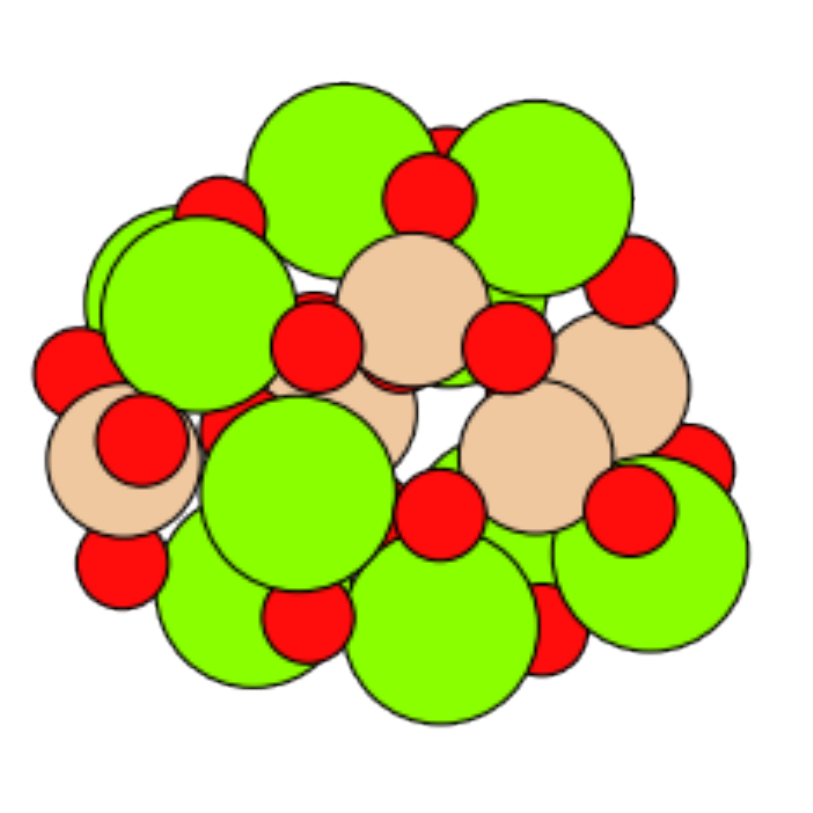}{5} & \fufig{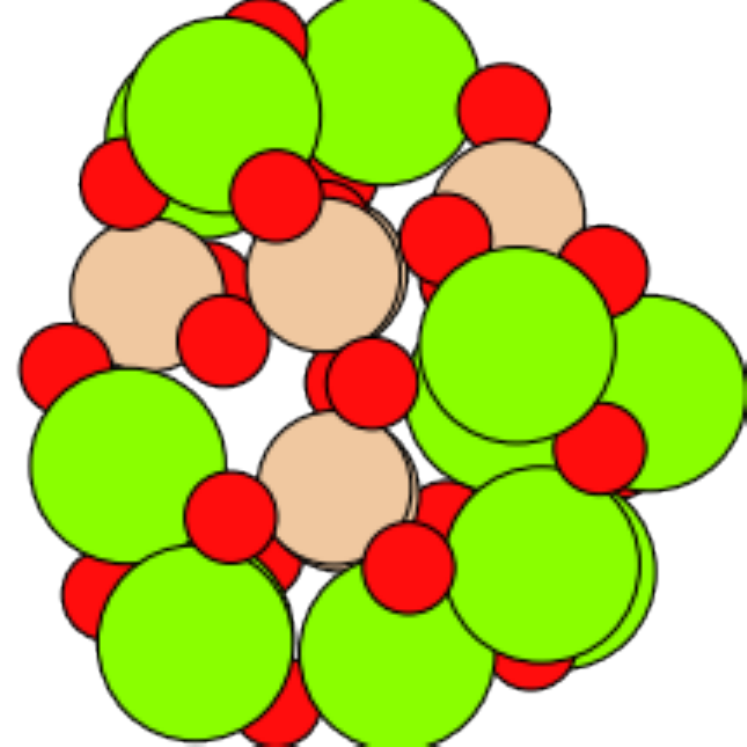}{6} & \fufig{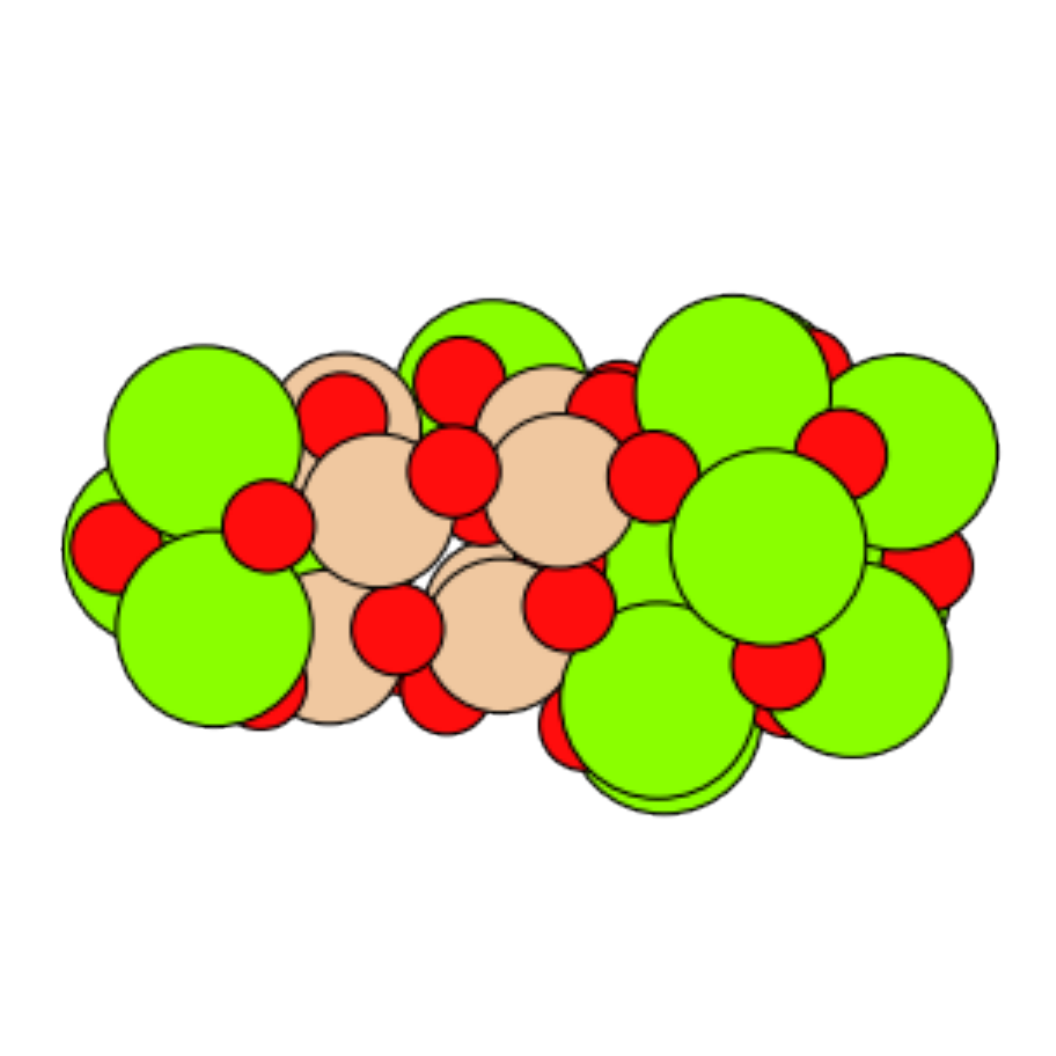}{7} \\
		\fufig{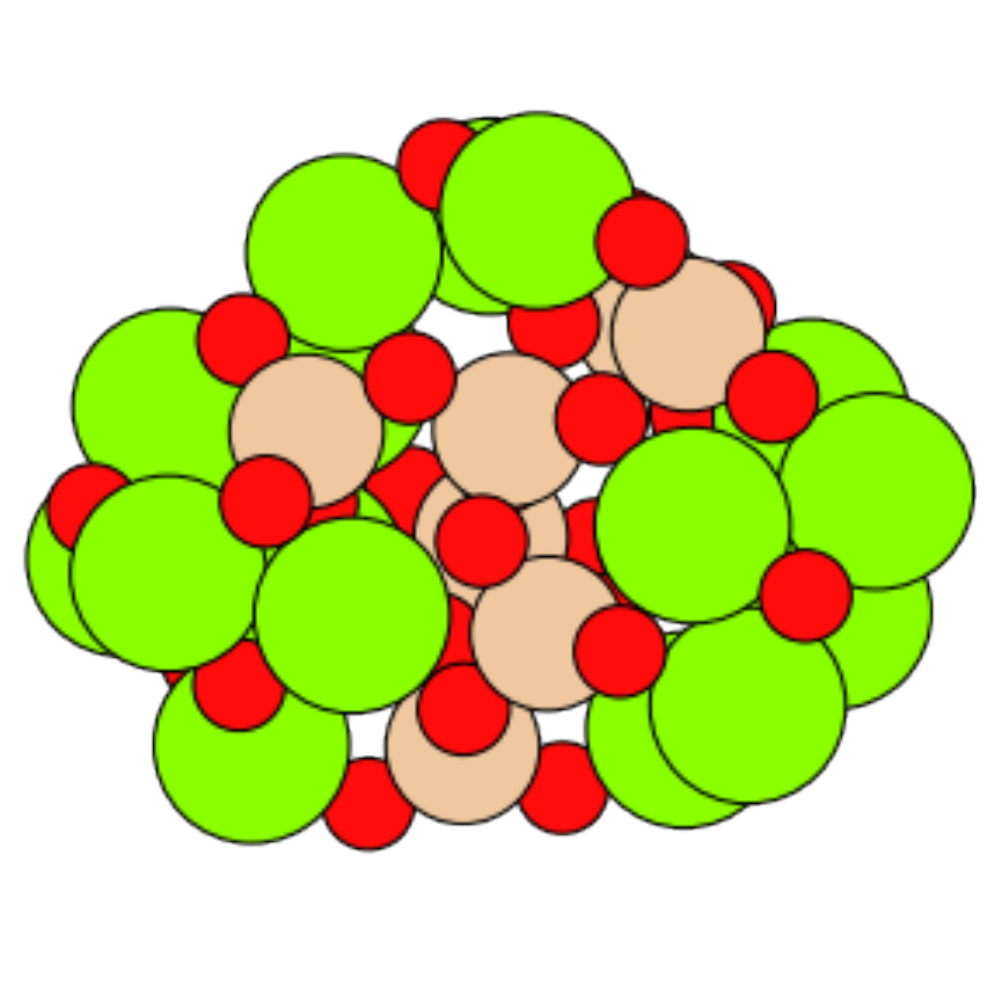}{8} & \fufig{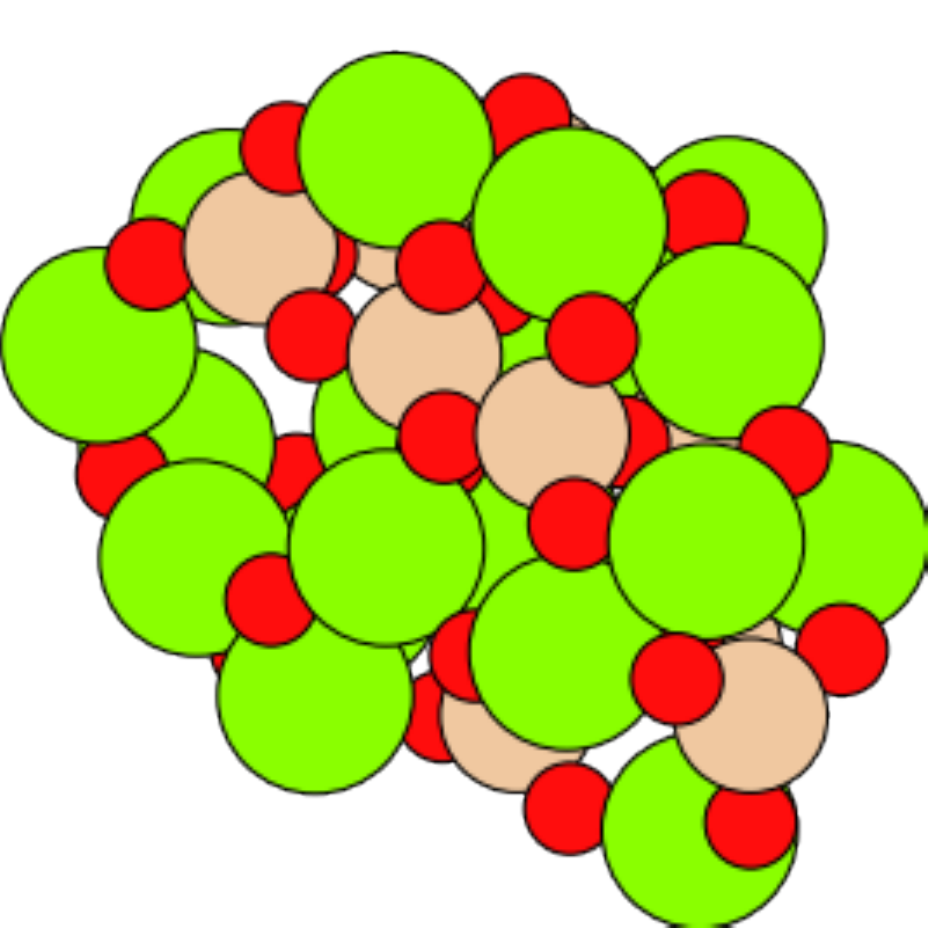}{9} & \fufig{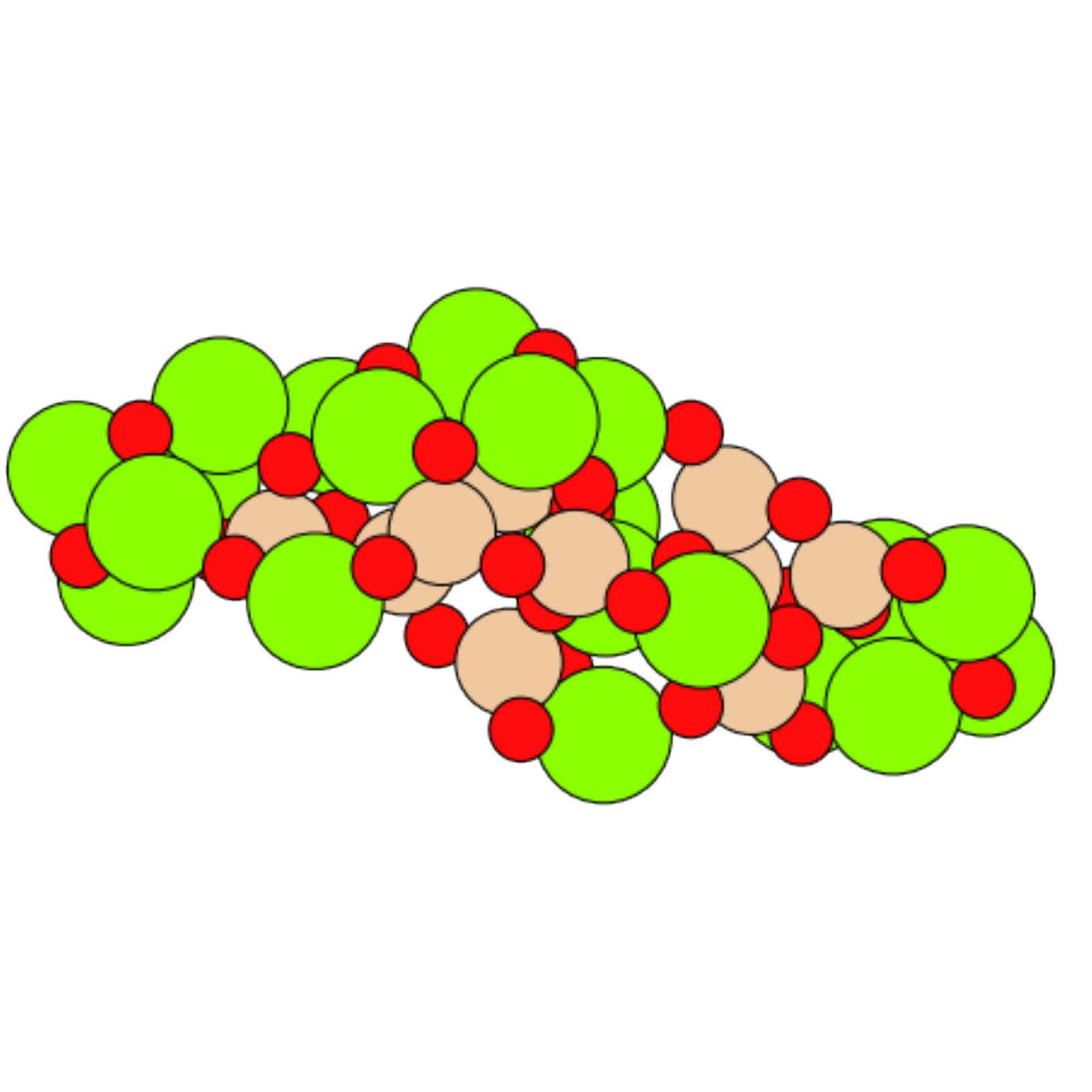}{10} \\
		\fufig{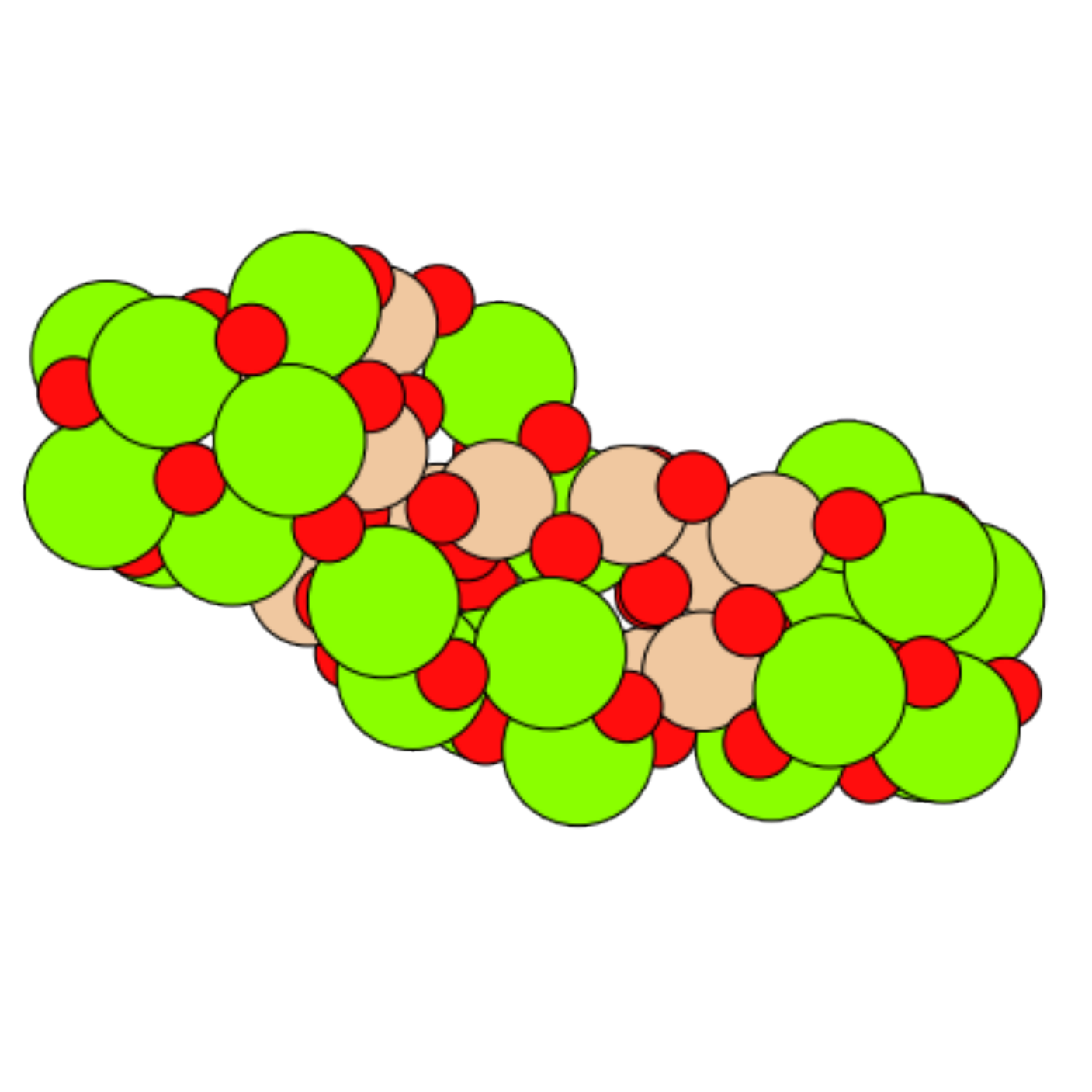}{11} & \fufig{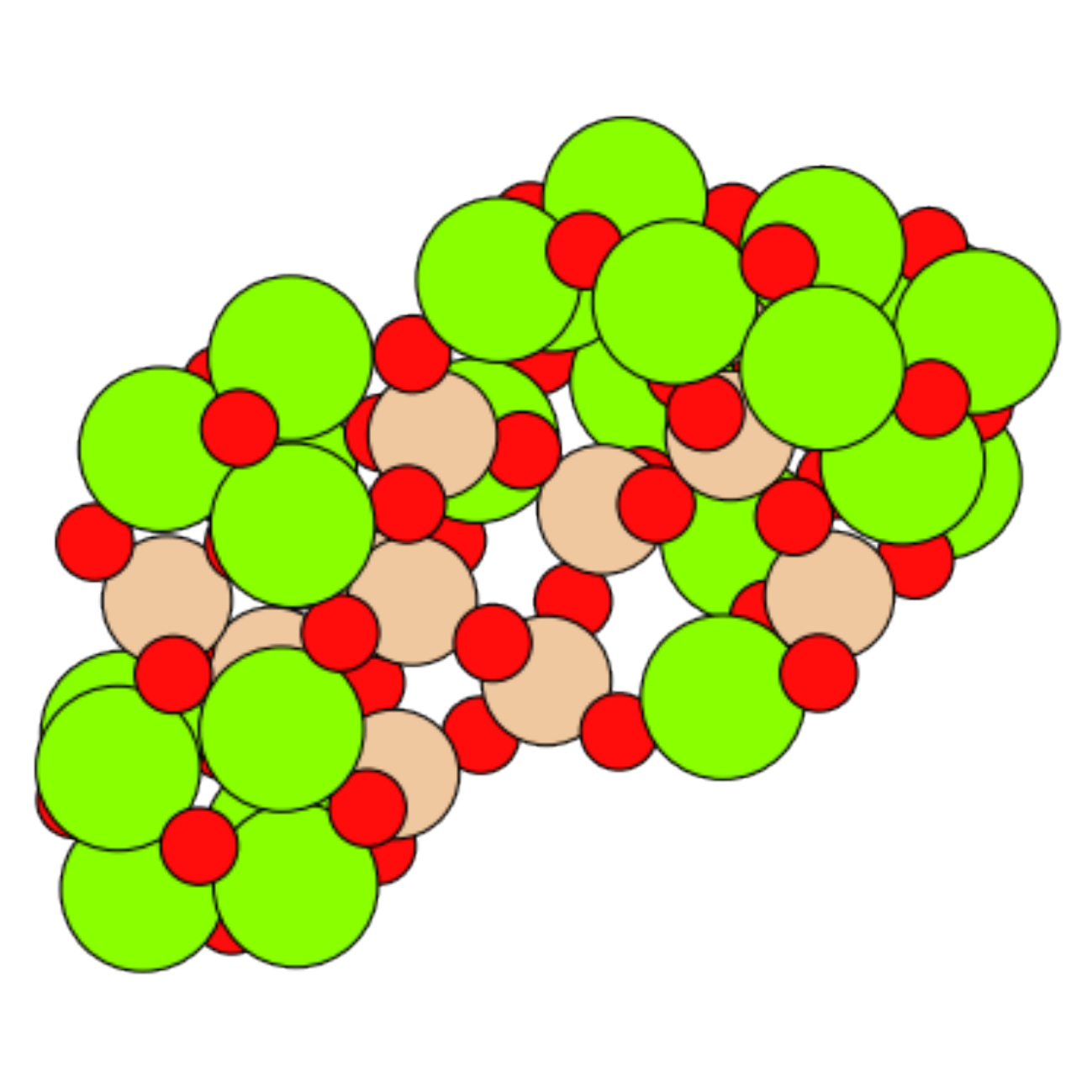}{12} & \fufig{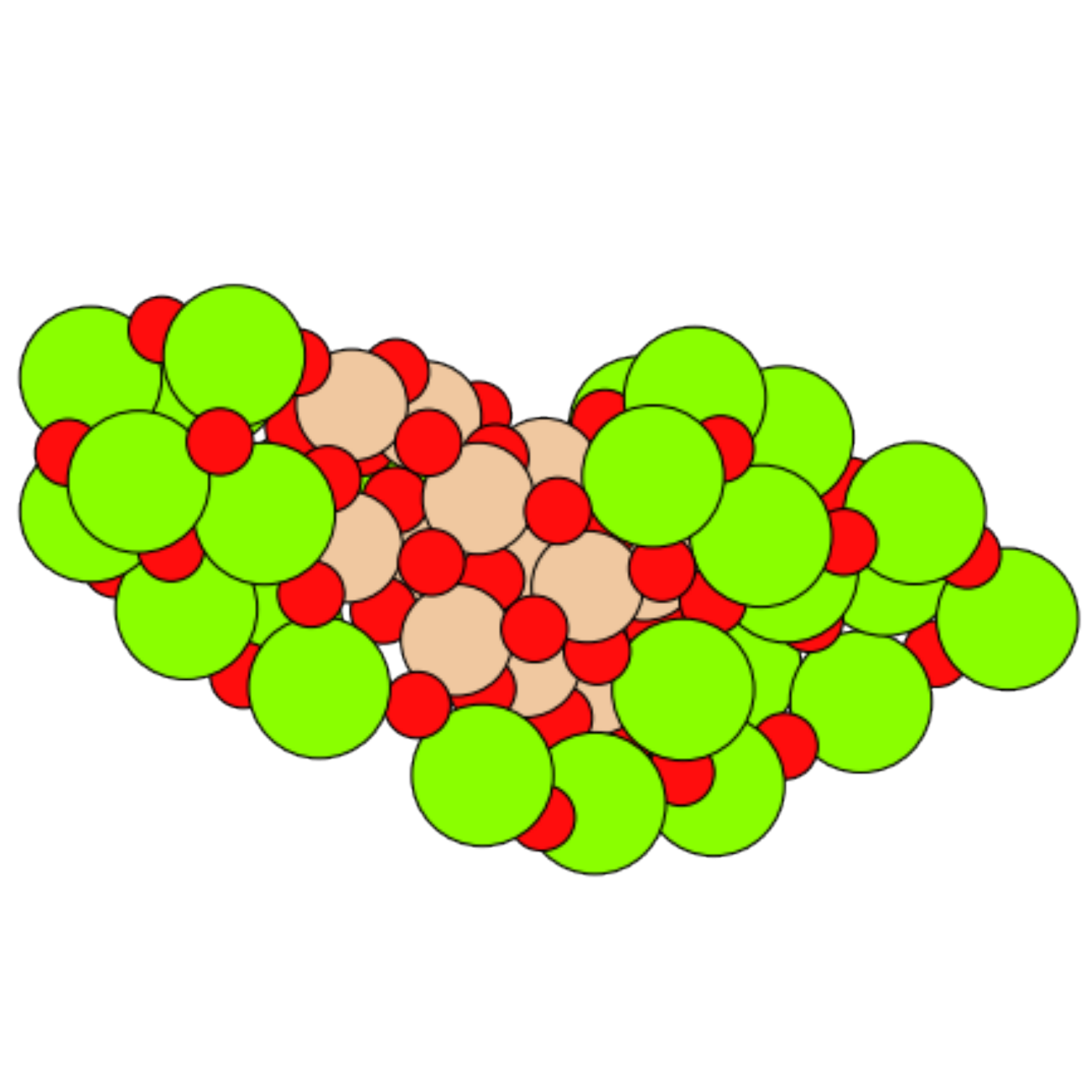}{13} \\
	\end{tabular}
	\caption{Clusters configurations found after DFT optimization. Colors represent atomic species: oxygen(red, small), silicon(tan, medium), magnesium(green, large).}
\label{fig:fu_table}
\end{figure}

%In a lattice, Fosterite will form with a regular orthorhombic structure. The crystal cell is composed of SiO$_4^4-$ and Mg$^{2+}$ in a 1:2 molar mass ratio. The empirical potential \ref{eq:buckingham_potential_0} is constructed such that the SiO$_4^4-$ anion is preferred during geometric relaxation.
Our clusters are shown in Fig. \ref{fig:fu_table}. Small molecules exhibit distinct symmetries, whereas large molecules become more amorphous, but with an underlying layering of MgO and SiO$_4$. We find no discernible tendency towards a bulk forsterite lattice structure, and expect clusters consisting of much larger numbers of monomers are necessary for a recognizable lattice structure to form. See the work of \citep{Horbach96, Noritake14} for a more systematic overview of size effects in silicate dynamics.

\begin{figure}[!ht]
	\centering
	\includegraphics[width=1.0\textwidth]{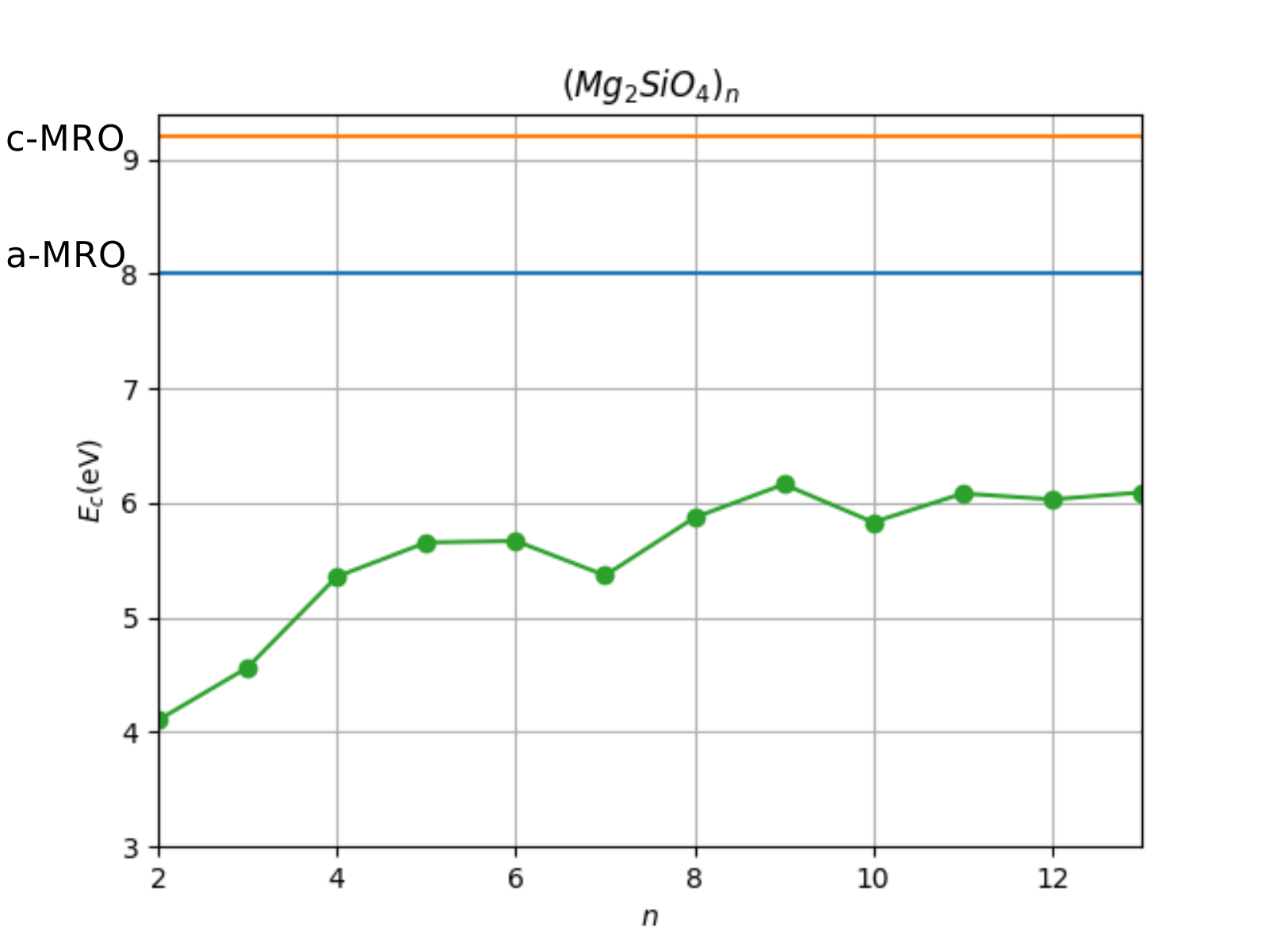}
	\caption{Cohesive energies (=$E_b/n$) of \fun{n} clusters. The constant plots c-MRO and a-MRO represent the bulk cohesive energy of crystal and amorphous \fun{n}, respectively.}
	\label{fig:ebnd_0}
\end{figure}

Cohesive energies ($E_b / n$) are plotted in Fig. (\ref{fig:ebnd_0}), along with the selected value of $\lambda$. These plots suggest convergence of these values at large monomer numbers. Edge and surface effects are still prominent in this regime (Horbach et all 1996), leading to non-monotonic growth of the cohesive energies.

%The coordination number per atom species is shown in Figure \ref{fig:coord_0}. \fun{7} has an above average number of bonds per atom, however bonding is not the dominate contribution to the binding energy.

%\begin{figure}[!ht]
%	\centering
%	\includegraphics[width=0.5\textwidth]{figures/coordination_0}
%	\caption{Coordination number of atomic species in Mg$_2$SiO$_4$)$_n$ clusters.}
%	\label{fig:coord_0}
%\end{figure}

\subsection{Critical sizes and nucleation rates}
\begin{figure}[!ht]
	\centering
	\includegraphics[width=1.0\textwidth]{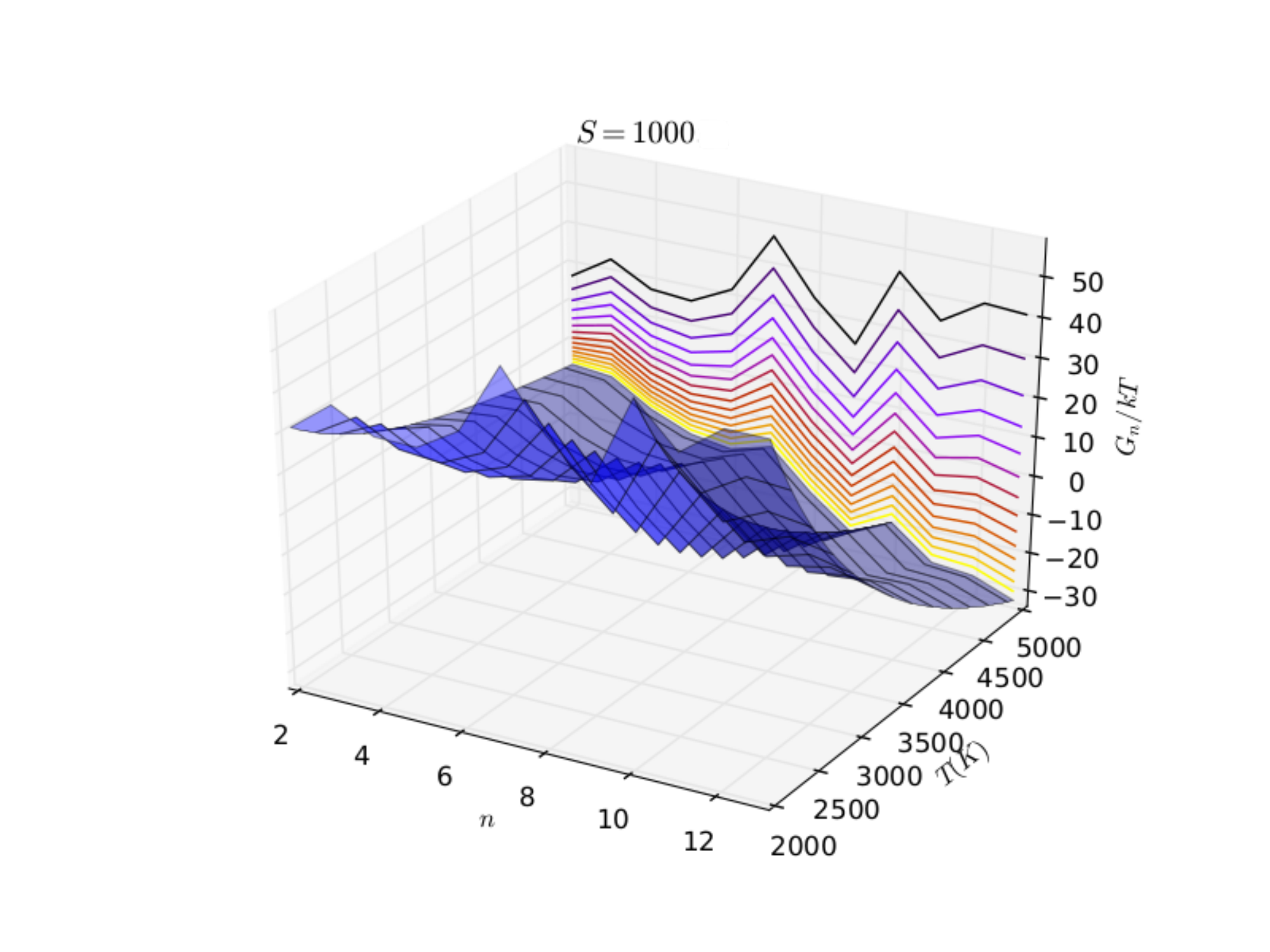}
	\caption{Work of Cluster Formation for $S=1000$ plotted across a range of temperatures. At high temperatures the curve is flattened and critical clusters will form at small $n$. For lower temperatures the WCF is more jagged, leading to higher critical sizes.}
	\label{fig:wcf_s1000}
\end{figure}

With binding energies at hand, we can find the free energy values and determine critical sizes and nucleation rates. Fig (\ref{fig:wcf_s1000}) is an example of free-energy curves at selected value of saturation across a range of temperatures. At low temperature values the WCF has no maximum in the range of monomer numbers, indicating a critical size will be larger than the $n*=13$, the largest cluster we studied. For the sake of completeness we would like to investigate clusters in the regime of $n>13$, these clusters are too large to be efficiently computed with our minima search and DFT calculations. Our size limit of $n=13$ is equivalent to 91 atoms, already stretching the limits of our computational techniques. Physically, moreover, molecules of this size have a very slow formation rate, and their contribution to dust creation can be taken to be negligible. Conversely, at large values of $(T,S)$ there is little to no free energy barrier and nucleation begins quickly with small clusters $n \approx 2$.

\begin{figure}[!ht]
	\centering
	\includegraphics[width=1.0\textwidth]{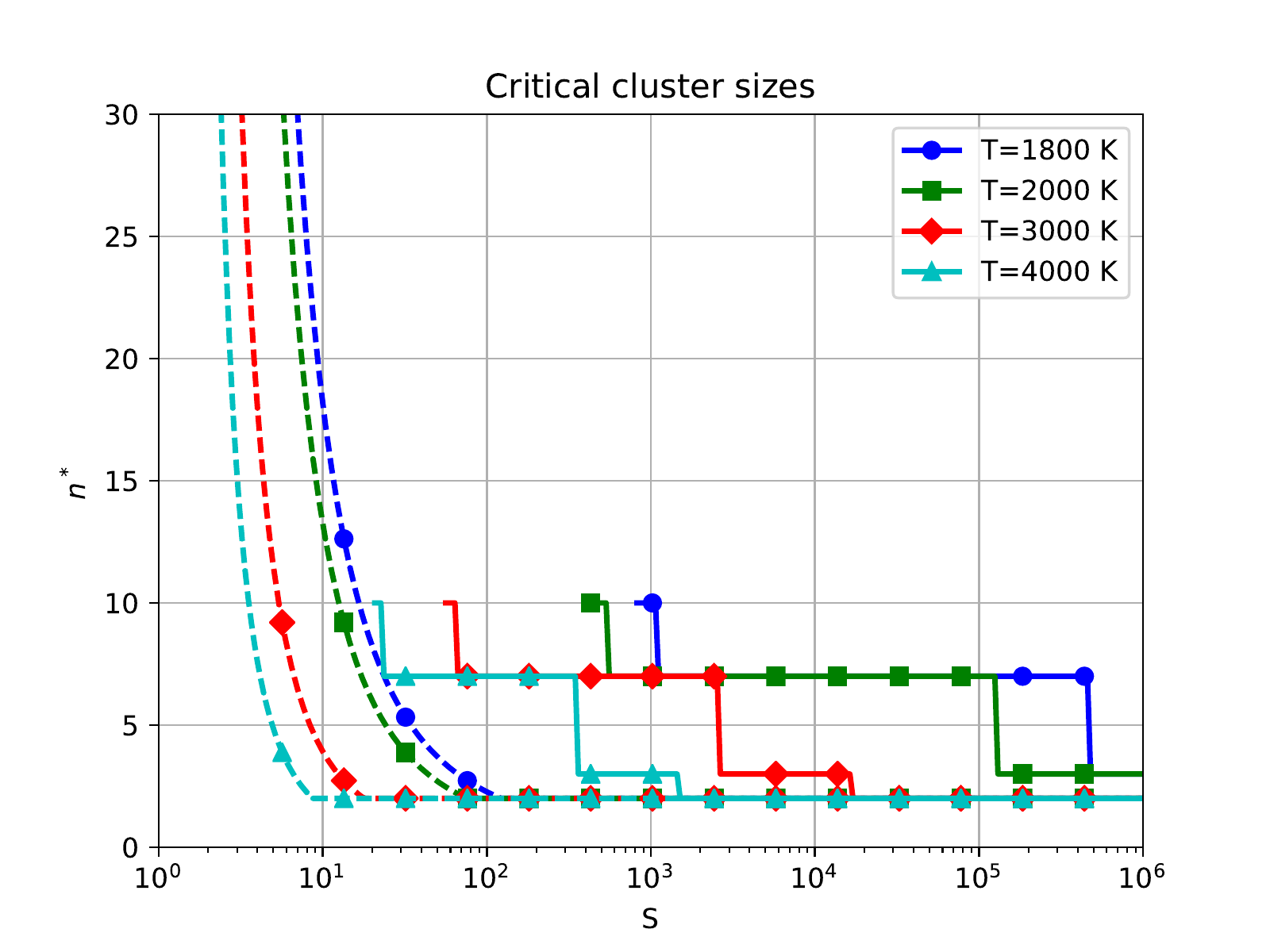}
	\caption{Critical size of \fun{n} clusters as a function of saturation at constant temperatures. Dashed lines are the CNT result, solid lines are the results of this article. Note that we impose a minimum critical size of $n=2$.}
	\label{fig:ncrit_l8-0}
\end{figure}

The WCF plots constructed, we can locate the maximum value and determine the critical cluster size $n^*$. These critical sizes across a range of temperatures and saturations are plotted in Fig. (\ref{fig:ncrit_l8-0}), along with results from CNT for comparison. As expected $n^*=2, 3$ for extreme environments, where density and temperature reduce the free energy barrier to the new phase and nucleation begins quickly. $n^*=7$ is prominent in the middle regions. Comparatively colder and sparser regions of the vapor have larger free energy barriers, leading to larger critical sizes $n* \ge 10$.

\begin{figure}[!ht]
	\centering
	\includegraphics[width=1.0\textwidth]{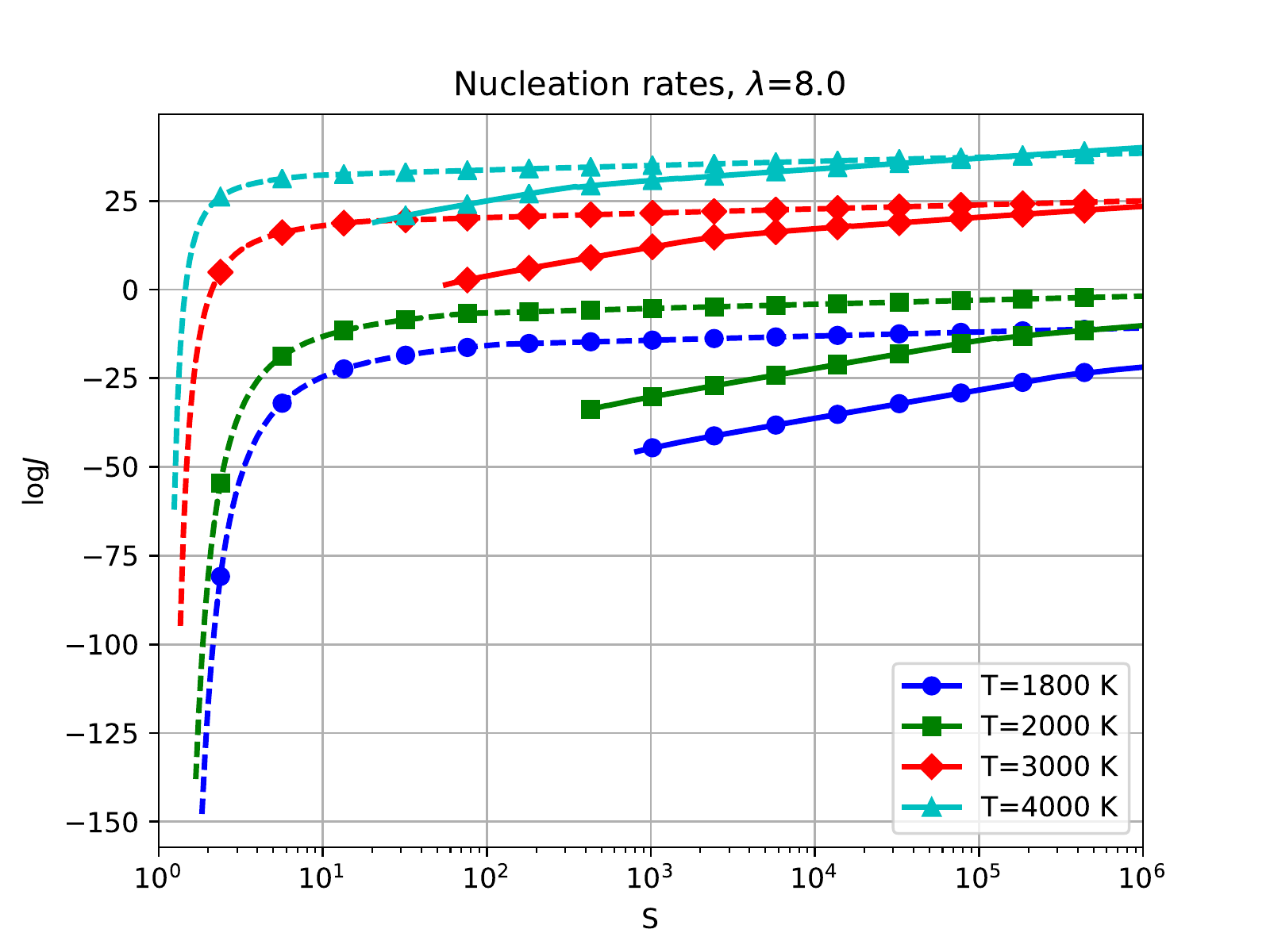}
	\caption{Nucleation rates of \fun{n} clusters as a function of saturation at constant temperatures. Dashed lines are the CNT result, solid lines are the results of this article.}
	\label{fig:jrate_l8-0}
\end{figure}

With critical sizes determined, nucleation rates follow from Eq. (\ref{eq:jrate}), and are plotted in Fig. (\ref{fig:jrate_l8-0}) along with the results from CNT for comparison. Nucleation is suppressed by several magnitudes at all but the largest temperatures and saturations compared to the classical case. Our results imply that nucleation does not take place at significant rates until at least several tens of saturation. This result is consistent with the critical sizes at low saturations being very large.

%No nucleation is present in environments of low saturation. Significant nucleation rates begin at large saturations. In Fig. (\ref{fig:jrate_ratio}) we compare our nucleation rates to those from CNT. Nucleation rates are suppressed compared to the classical theory at all values of temperature and saturation.

% \begin{figure}[!ht]
% 	\centering
% 	\includegraphics[width=0.5\textwidth]{figures/jrate_ratio}
% 	\caption{Comparison with classical nucleation rates. $J/J_C$}
% 	\label{fig:jrate_ratio}
% \end{figure}
 
\section{Conclusions and Discussion}
\label{sec:conclusion}
We find new results for the ground state configuration of \fun{n} clusters and determine the rates of nucleation of MRO clusters in stellar outflows. Using DFT we have determined precise values of binding energies for these clusters, and used these binding energies in an atomistic formulation of nucleation theory to produce nucleation rates for silicates. These values have been compared to CNT and have been found to be significantly different than the classical case. Except for environments of large temperatures and saturations, out ANT approach finds lower formation rates of critical clusters when compared to CNT.

While we expect this trend to hold for regions of low temperature and saturations, the critical sizes that would be expected to be found at these environments are too large for our methods to efficiently determine. It is possible that at low temperatures and at saturations of approximately unity, ANT will predict enhanced nucleation over that of CNT. However, even in CNT nucleation in these environments is negligibly small (see, for instance, the low saturation regions of Figure \ref{fig:jrate_l8-0}).

Locating the ground state of large multi-component systems is difficult. Further limiting research into these clusters is the use of the empirical BKS model in studying nanoscale clusters. This potential form is known to perform poorly when describing the surface chemistry of silicates\citep{ceresoli2000two,Flikkeme03}. Ideally an \emph{ab-initio} ground-state search would be preferable. However, global minima techniques require many million energy evaluations, and at present this approach would too computationally prohibitive.

Except for the small clusters $n=2$, $n=3$, and $n=4$ the cluster configurations found do not exhibit strong symmetries or growth patterns. Larger clusters are amorphous and lack any well-structured ground state. This can be seen in the binding energy of the clusters (Figure \ref{fig:ebnd_0}), which stops growing monotonically. In the specific cases of $n=7$ and $n=10$ there is a significant drop in stability. While it is possible that there are configurations of clusters of comparable energies, the relative energies of the clusters will remain the same (for instance, $n=7$ and $n=10$ showing lowered stability.)

\begin{figure}[!ht]
	\centering
	\includegraphics[width=1.0\textwidth]{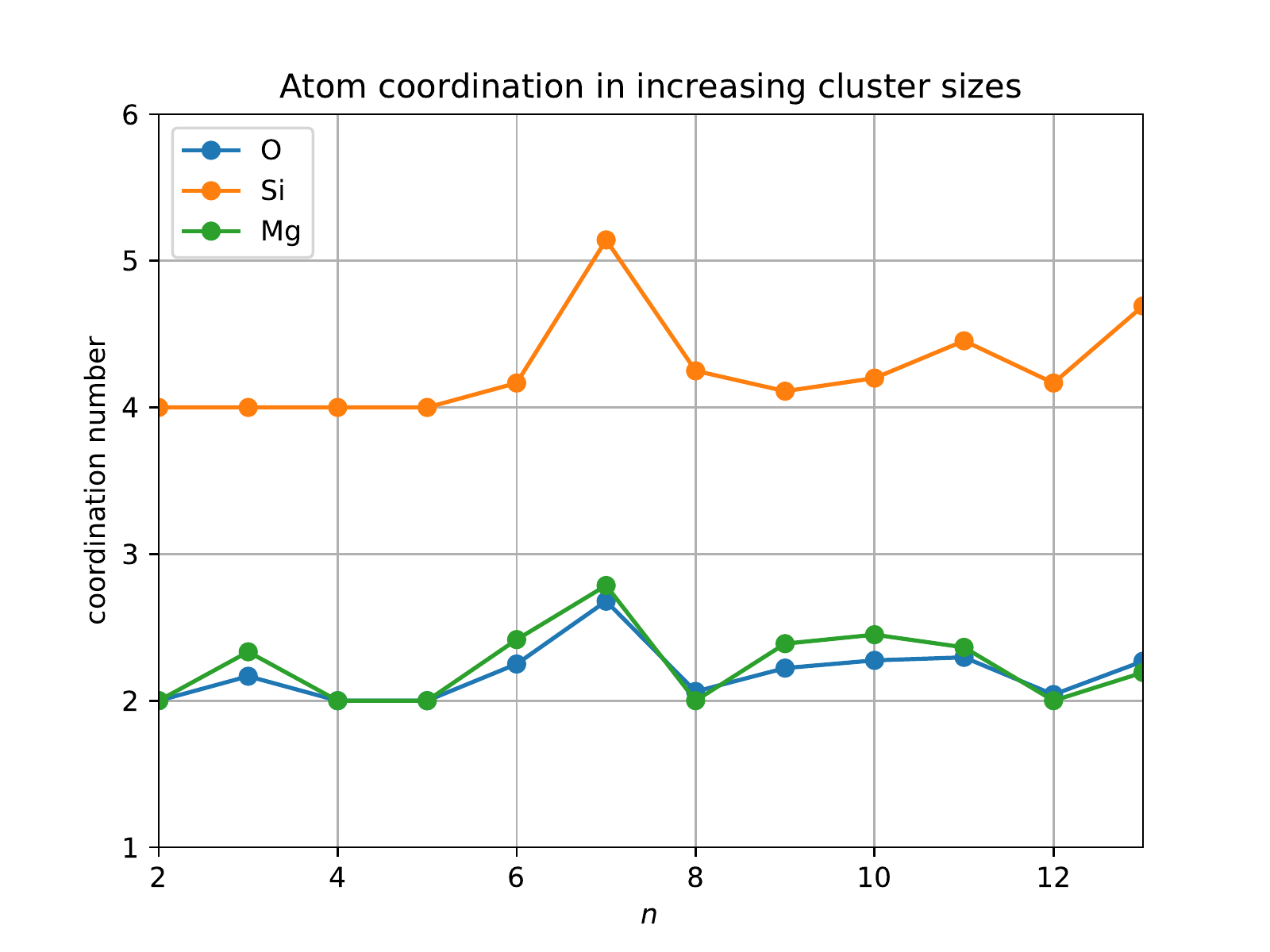}
	\caption{Average coordination of each atomic species plotted against cluster size. Increased Si coordination is seen at $n=7$ and $n=10$.}
	\label{fig:coordination_0}
\end{figure}

The drop in stability at $n=7$ and $n=10$ can be explained by examining the mean coordination (number of bonds) in these clusters. As seen in Figure \ref{fig:coordination_0}, there is an increase in Si coordination at $n=7$, indicating that the Si atoms are being weakly bonded to fifth O atom and weakening the stability of the SiO$_4$ tetrahedra. A similar, albeit smaller, effect is also noticeable at $n=10$. It is not surprising, then, that these clusters turn out to be prominent critical sizes. Adding a monomer from the vapor to these clusters will release more free energy by lowering the mean coordination of the SiO$_4$ tetrahedra.  At larger cluster sizes there is an overall trend in increasing Si coordination as a result of MgO-SiO$_4$ layering (see below).

%This is likely the result of missing a more stable configuration in the MH algorithm. This leads to a bias in the critical size results for these clusters. However, because the differences in cohesive energies of clusters of these sizes is small, the nucleation rates remain largely unaltered.
%This is due to several reasons. First is the difficulty in determining the ground state of large, multi-component systems, which occupy a staggeringly large phase space, and efficient minima searches of such systems are largely intractable. Few studies exist that identify highly symmetric ground states of large molecules consisting of three or more constituent atoms. The second challenge is using the empirical BKS model in studying nanoscale clusters.  This potential form is known to perform poorly when describing the surface chemistry of silicates\citep{ceresoli2000two,Flikkeme03}. Ideally an \emph{ab-initio} ground-state search would be preferable, although without significant computing power this is not feasible. 

The layering present in our results is explained by Noritake \citep{Noritake14}. Briefly, Si-O bonds are much stronger than Mg-O bonds, leading to Si atoms preferentially sharing an oxygen bond as Si-O-Si. Thus SiO$_4$ tetrahedra cluster together and have overlapping O sites. These shared O atoms are not available to bond to Mg atoms, and the Mg will gather around excess O atoms. Our results suggest that during nucleation MROs begin as a melt, and that the emergence of Mg-Si layering explains the location of MRO critical cluster sizes.

Our results support the model that MRO dust precursor molecules form amorphously, when following a fixed stoichiometry formation path. Evidence from simulations of silicate glasses \citep{Horbach96} suggests that the transition from silicate melt to crystal lattice occurs in ensembles of several hundred to thousands of atoms. Crystalline silicates in late-stage stellar environments will therefore form after a period of processing and annealing.  The presence of OH and Fe in the surrounding vapor will enhance the formation of crystalline MROs \citep{jager1998steps}, and their inclusion in nucleation studies may result in more crystalline-like ground states.

Critical sizes for MRO clusters are large in regions of low pressure and temperature. In Fig. \ref{fig:ncrit_l8-0} it can be seen that regions of low temperature and pressure show no critical size in the range of sizes we considered. In the absence of a critical size it is not possible to determine the nucleation rates in ANT. However, as can be seen in Fig. \ref{fig:jrate_l8-0}, nucleation rates in these regions will be near negligible, and nucleation in this regime will not be a significant contributor to dust creation. This does not mean that MRO nucleation at low temperature is impossible. Rather, the gas must achieve a very large saturation before nucleation of MRO clusters becomes efficient. At lower temperatures, nucleation of smaller silicate clusters (MgO, SiO) is more prominent \citep{Jeong03,Goumans12}. However, at this stage most of the condensable monomers will have nucleated into MRO clusters. 

Nucleation rates are suppressed compared to CNT for all conditions save for the largest saturations we studied. In the classical theory, the value of the surface tension allows for clusters to quickly grow to the critical cluster size and begin nucleation. However, our results indicate that this value of surface tension is too low in the classical case (for instance the value given in \citep{Boni56} $\sigma=425$ erg cm$^-2$), and does not accurately represent the surface physics of nanoscale silicate clusters.

Silicate nucleation rates are suppressed at low temperatures and saturations. This suggests 'staged' nucleation periods, with carbon forming and growing quickly while silicate clusters take longer to form. Interestingly, the chemical kinetics approach reverses the priority of this staging, with carbon dust forming later than silicate clusters \citep{Sarangi15}. More detailed observations of the dust formation histories of CCNSe outflows will be useful in resolving this disagreement between models. 

This work was supported in part by NSF grants
1150365-AST and 1461362-AST (DL \& CM). We also would like to thank Professors Goumans and Bromley for sharing their research and providing insights into silicate cluster growth.

%\bibliographystyle{apalike}
%\bibliography{bib.bib}

%\bibliographystyle{plain}
\bibliographystyle{elsarticle-num}
\bibliography{bib_elsevier}

\begin{thebibliography}{10}
\expandafter\ifx\csname url\endcsname\relax
  \def\url#1{\texttt{#1}}\fi
\expandafter\ifx\csname urlprefix\endcsname\relax\def\urlprefix{URL }\fi
\expandafter\ifx\csname href\endcsname\relax
  \def\href#1#2{#2} \def\path#1{#1}\fi

\bibitem{gibb2004interstellar}
E.~Gibb, D.~Whittet, A.~Boogert, A.~Tielens, Interstellar ice: the infrared
  space observatory legacy, The Astrophysical Journal Supplement Series 151~(1)
  (2004) 35.

\bibitem{teplitz2006silicate}
H.~I. Teplitz, L.~Armus, B.~Soifer, V.~Charmandaris, J.~Marshall, H.~Spoon,
  C.~Lawrence, L.~Hao, S.~Higdon, Y.~Wu, et~al., Silicate emission in the
  spitzer irs spectrum of fsc 10214+ 4724, The Astrophysical Journal Letters
  638~(1) (2006) L1.

\bibitem{roche2007silicate}
P.~F. Roche, C.~Packham, D.~K. Aitken, R.~E. Mason, Silicate absorption in
  heavily obscured galaxy nuclei, Monthly Notices of the Royal Astronomical
  Society 375~(1) (2007) 99--104.

\bibitem{kemper2004absence}
F.~Kemper, W.~Vriend, A.~Tielens, The absence of crystalline silicates in the
  diffuse interstellar medium, The Astrophysical Journal 609~(2) (2004) 826.

\bibitem{molster2002crystalline}
F.~Molster, L.~Waters, A.~Tielens, M.~Barlow, Crystalline silicate dust around
  evolved stars-i. the sample stars, Astronomy \& Astrophysics 382~(1) (2002)
  184--221.

\bibitem{olofsson2009c2d}
J.~Olofsson, J.-C. Augereau, E.~Van~Dishoeck, B.~Mer{\'\i}n, F.~Lahuis,
  J.~Kessler-Silacci, C.~Dullemond, I.~Oliveira, G.~Blake, A.~Boogert, et~al.,
  C2d spitzer-irs spectra of disks around t tauri stars-iv. crystalline
  silicates, Astronomy \& Astrophysics 507~(1) (2009) 327--345.

\bibitem{wooden1999silicate}
D.~H. Wooden, D.~E. Harker, C.~E. Woodward, H.~M. Butner, C.~Koike, F.~C.
  Witteborn, C.~W. McMurtry, Silicate mineralogy of the dust in the inner coma
  of comet c/1995 01 (hale-bopp) pre-and postperihelion, The Astrophysical
  Journal 517~(2) (1999) 1034.

\bibitem{henning2010cosmic}
T.~Henning, Cosmic silicates, Annual Review of Astronomy and Astrophysics 48
  (2010) 21--46.

\bibitem{Mauney15}
C.~Mauney, M.~Buongiorno~Nardelli, D.~Lazzati, The Astrophysical Journal
  800~(30).

\bibitem{Todini01}
P.~{Todini}, A.~{Ferrara}, {Dust formation in primordial Type II supernovae},
  Monthly Notices of the RAS 325 (2001) 726--736.
\newblock \href {http://arxiv.org/abs/astro-ph/0009176}
  {\path{arXiv:astro-ph/0009176}}, \href
  {http://dx.doi.org/10.1046/j.1365-8711.2001.04486.x}
  {\path{doi:10.1046/j.1365-8711.2001.04486.x}}.

\bibitem{Nozawa03}
T.~{Nozawa}, T.~{Kozasa}, H.~{Umeda}, K.~{Maeda}, K.~{Nomoto}, {Dust in the
  Early Universe: Dust Formation in the Ejecta of Population III Supernovae},
  Astrophysical Journal 598 (2003) 785--803.
\newblock \href {http://arxiv.org/abs/astro-ph/0307108}
  {\path{arXiv:astro-ph/0307108}}, \href {http://dx.doi.org/10.1086/379011}
  {\path{doi:10.1086/379011}}.

\bibitem{Goumans12}
T.~P.~M. {Goumans}, S.~T. {Bromley}, {Efficient nucleation of stardust
  silicates via heteromolecular homogeneous condensation}, Monthly Notices of
  the RAS 420 (2012) 3344--3349.
\newblock \href {http://dx.doi.org/10.1111/j.1365-2966.2011.20255.x}
  {\path{doi:10.1111/j.1365-2966.2011.20255.x}}.

\bibitem{Bromley16}
S.~T. Bromley, J.~C. Gomez~Martin, J.~M.~C. Plane,
  \href{http://dx.doi.org/10.1039/C6CP03629E}{Under what conditions does (sio)n
  nucleation occur? a bottom-up kinetic modelling evaluation}, Phys. Chem.
  Chem. Phys. 18 (2016) 26913--26922.
\newblock \href {http://dx.doi.org/10.1039/C6CP03629E}
  {\path{doi:10.1039/C6CP03629E}}.
\newline\urlprefix\url{http://dx.doi.org/10.1039/C6CP03629E}

\bibitem{Goedecker04}
S.~Goedecker, Journal of Chemical Physics 120~(21).

\bibitem{rondina2013revised}
G.~G. Rondina, J.~L. Da~Silva, Revised basin-hopping monte carlo algorithm for
  structure optimization of clusters and nanoparticles, Journal of chemical
  information and modeling 53~(9) (2013) 2282--2298.

\bibitem{van1990force}
B.~Van~Beest, G.~J. Kramer, R.~Van~Santen, Force fields for silicas and
  aluminophosphates based on ab initio calculations, Physical Review Letters
  64~(16) (1990) 1955.

\bibitem{hassanali2007model}
A.~A. Hassanali, S.~J. Singer, Model for the water- amorphous silica interface:
  The undissociated surface, The Journal of Physical Chemistry B 111~(38)
  (2007) 11181--11193.

\bibitem{roberts2001investigation}
C.~Roberts, R.~L. Johnston, Investigation of the structures of mgo clusters
  using a genetic algorithm, Physical Chemistry Chemical Physics 3~(22) (2001)
  5024--5034.

\bibitem{flikkema2003new}
E.~Flikkema, S.~T. Bromley, A new interatomic potential for nanoscale silica,
  Chemical Physics Letters 378~(5) (2003) 622--629.

\bibitem{shanno1985broyden}
D.~F. Shanno, On broyden-fletcher-goldfarb-shanno method, Journal of
  Optimization Theory and Applications 46~(1) (1985) 87--94.

\bibitem{Deaven95}
D.~M. {Deaven}, K.~M. {Ho}, {Molecular Geometry Optimization with a Genetic
  Algorithm}, Physical Review Letters 75 (1995) 288--291.
\newblock \href {http://dx.doi.org/10.1103/PhysRevLett.75.288}
  {\path{doi:10.1103/PhysRevLett.75.288}}.

\bibitem{giannozzi2009quantum}
P.~Giannozzi, S.~Baroni, N.~Bonini, M.~Calandra, R.~Car, C.~Cavazzoni,
  D.~Ceresoli, G.~L. Chiarotti, M.~Cococcioni, I.~Dabo, et~al., Quantum
  espresso: a modular and open-source software project for quantum simulations
  of materials, Journal of physics: Condensed matter 21~(39) (2009) 395502.

\bibitem{MartinTB}
R.~M. Martin, Electronic Structure, Cambridge University Press, 2004.

\bibitem{Kozasa87}
T.~{Kozasa}, H.~{Hasegawa}, {Grain Formation through Nucleation Process in
  Astrophysical Environments. II ---Nucleation and Grain Growth Accompanied by
  Chemical Reaction---}, Progress of Theoretical Physics 77 (1987) 1402--1410.
\newblock \href {http://dx.doi.org/10.1143/PTP.77.1402}
  {\path{doi:10.1143/PTP.77.1402}}.

\bibitem{thogersen2008experimental}
A.~Th{\o}gersen, S.~Diplas, J.~Mayandi, T.~Finstad, A.~Olsen, J.~F. Watts,
  M.~Mitome, Y.~Bando, An experimental study of charge distribution in
  crystalline and amorphous si nanoclusters in thin silica films, Journal of
  Applied Physics 103~(2) (2008) 024308.

\bibitem{gonccalves2016molecular}
W.~Gon{\c{c}}alves, J.~Morthomas, P.~Chantrenne, M.~Perez, G.~Foray, C.~L.
  Martin, Molecular dynamics simulations of amorphous silica surface properties
  with truncated coulomb interactions, Journal of Non-Crystalline Solids 447
  (2016) 1--8.

\bibitem{Catti81}
M.~{Catti}, {The lattice energy of forsterite. Charge distribution and
  formation enthalpy of the SiO$\{$$_{4}$/$^{4-}$$\}$ ion}, Physics and
  Chemistry of Minerals 7 (1981) 20--25.
\newblock \href {http://dx.doi.org/10.1007/BF00308196}
  {\path{doi:10.1007/BF00308196}}.

\bibitem{May00}
P.~W. {May}, G.~{Pineau des For{\^e}ts}, D.~R. {Flower}, D.~{Field}, N.~L.
  {Allan}, J.~A. {Purton}, {Sputtering of grains in C-type shocks}, Monthly
  Notices of the RAS 318 (2000) 809--816.
\newblock \href {http://dx.doi.org/10.1046/j.1365-8711.2000.03796.x}
  {\path{doi:10.1046/j.1365-8711.2000.03796.x}}.

\bibitem{Horbach96}
J.~{Horbach}, W.~{Kob}, K.~{Binder}, C.~A. {Angell}, {Finite size effects in
  simulations of glass dynamics}, Physical Review E 54 (1996) R5897--R5900.
\newblock \href {http://arxiv.org/abs/cond-mat/9610066}
  {\path{arXiv:cond-mat/9610066}}, \href
  {http://dx.doi.org/10.1103/PhysRevE.54.R5897}
  {\path{doi:10.1103/PhysRevE.54.R5897}}.

\bibitem{Noritake14}
F.~{Noritake}, K.~{Kawamura}, {Structure and properties of
  forsterite-MgSiO$_{3}$ liquid interface: molecular dynamics study}, Progress
  in Earth and Planetary Science 1 (2014) 14.
\newblock \href {http://dx.doi.org/10.1186/2197-4284-1-14}
  {\path{doi:10.1186/2197-4284-1-14}}.

\bibitem{ceresoli2000two}
D.~Ceresoli, M.~Bernasconi, S.~Iarlori, M.~Parrinello, E.~Tosatti, Two-membered
  silicon rings on the dehydroxylated surface of silica, Physical review
  letters 84~(17) (2000) 3887.

\bibitem{Flikkeme03}
E.~Flikkema, S.~Bromley, Chemical Physics Letters 378~(5).

\bibitem{jager1998steps}
C.~J{\"a}ger, F.~Molster, J.~Dorschner, T.~Henning, H.~Mutschke, L.~Waters,
  Steps toward interstellar silicate mineralogy. iv. the crystalline
  revolution, Astronomy and Astrophysics 339 (1998) 904--916.

\bibitem{Jeong03}
K.~S. {Jeong}, J.~M. {Winters}, T.~{Le Bertre}, E.~{Sedlmayr}, {Self-consistent
  modeling of the outflow from the O-rich Mira IRC -20197}, Astronomy and
  Astrophysics 407 (2003) 191--206.
\newblock \href {http://dx.doi.org/10.1051/0004-6361:20030693}
  {\path{doi:10.1051/0004-6361:20030693}}.

\bibitem{Boni56}
R.~E. {Boni}, G.~{Derge}, {Surface Tensions of Silicates}, JOM - Journal of the
  Minerals, Metals and Materials Society 8 (1956) 53--59.
\newblock \href {http://dx.doi.org/10.1007/BF03377643}
  {\path{doi:10.1007/BF03377643}}.

\bibitem{Sarangi15}
A.~Sarangi, I.~Cherchneff, Astronomy and Astrophysics 575~(A95).

\end{thebibliography}

\appendix
\section{Effect of $\lambda$ values on nucleation rates}
The parameter $\lambda$ represents the work required to remove a monomer from the bulk solid phase. This value is used in the atomistic formulation of nucleation rates to determine the surface energy of a cluster, as in Eq. (\ref{eqn:gfe_S}). The complex nature of amorphous silicate growth prevents simple evaluation of this parameter. In section \ref{sec:methods} we present our approach to selection of this value for this work. In this appendix we present results of nucleation rates with lower and higher values of $\lambda$.

\begin{figure}[!ht]
	\centering
	\includegraphics[width=1.0\textwidth]{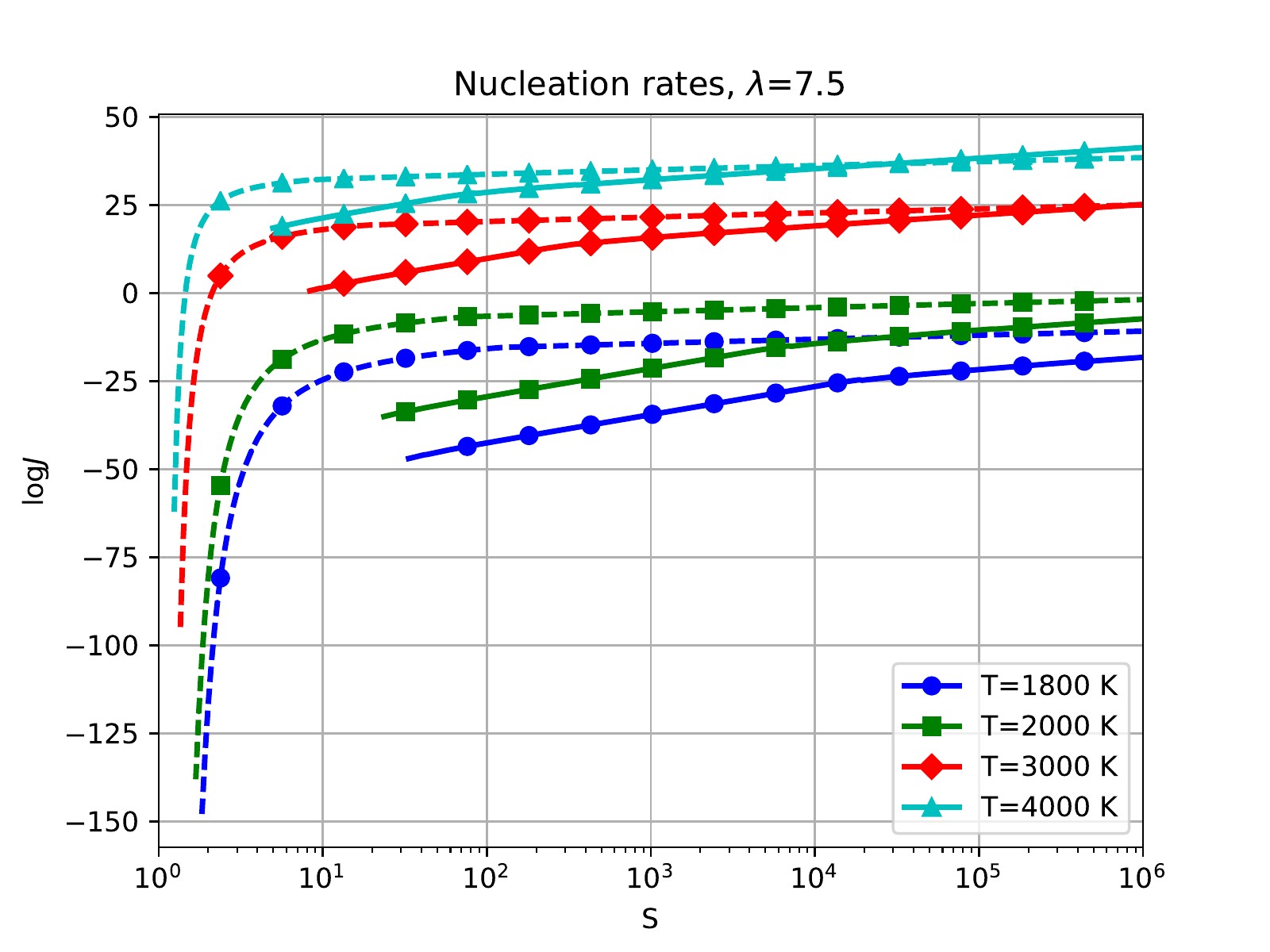}
	\caption{Nucleation rates at $\lambda = 7.5$ eV. Dashed lines are the CNT result, solid lines are the results of this article. Temperatures given in Kelvins.}
	\label{fig:jrate_l7-5}
\end{figure}

Lower values of $\lambda$ imply a lower energy barrier between phases, and this will induce faster nucleation. This can be seen in Fig. \ref{fig:jrate_l7-5}. Nucleation rates are still suppressed compared to classical nucleation, but lower saturations are required to begin nucleation.  The corresponds physically to a mostly free formation pathway. While our results indicate that silicate growth is amorphous, there is indeed some underlying constraints on growth (for instance, the MgO-SiO$_4$ layering) that would argue against using such a low value for $\lambda$.

\begin{figure}[!ht]
	\centering
	\includegraphics[width=1.0\textwidth]{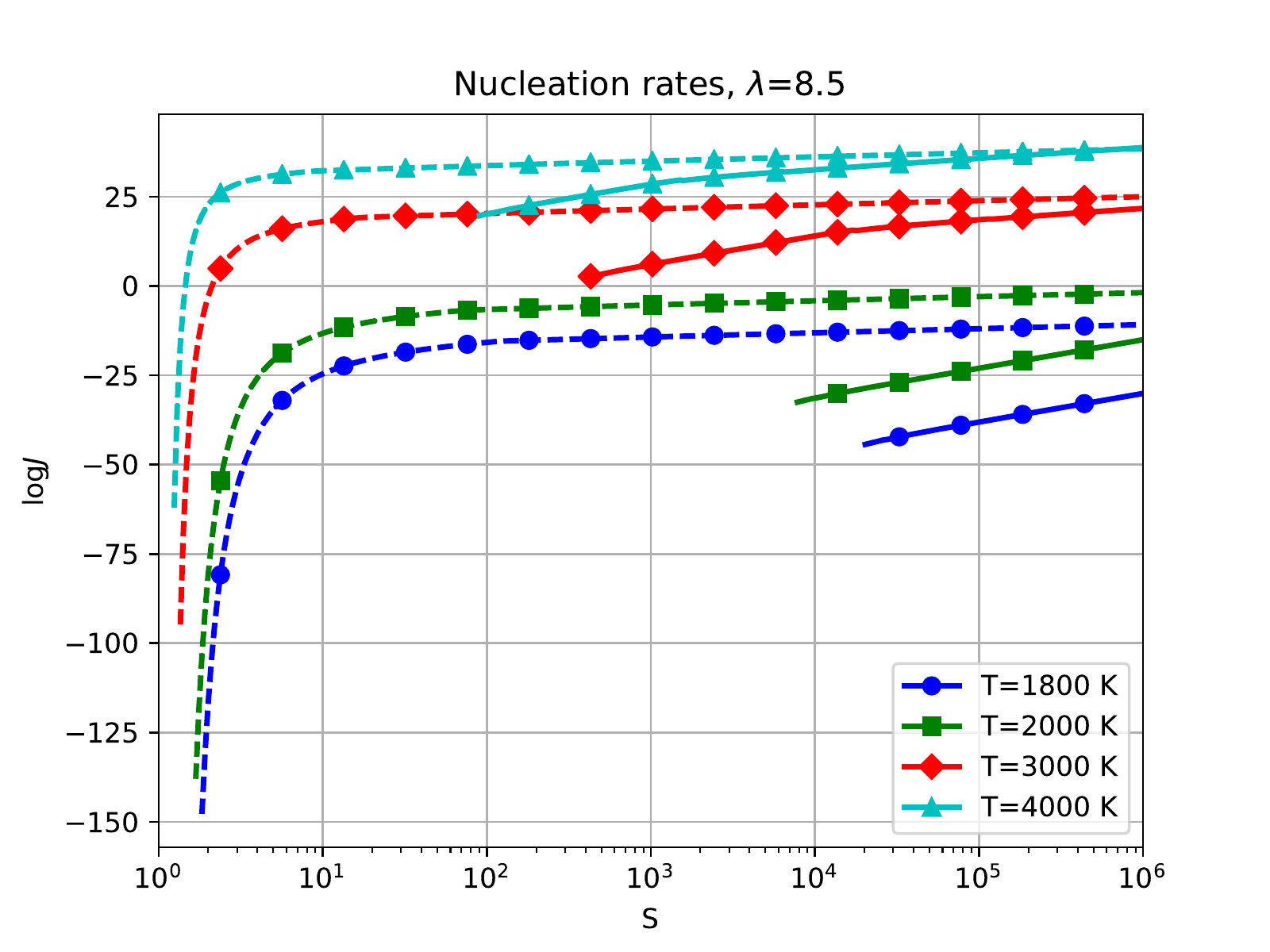}
	\caption{Nucleation rates at $\lambda = 8.5$ eV. Dashed lines are the CNT result, solid lines are the results of this article. Temperatures given in Kelvins.}
	\label{fig:jrate_l8-5}
\end{figure}

Fig. \ref{fig:jrate_l8-5} represents are larger value of $\lambda$, and shows significantly less nucleation. This corresponds to a much more constrained formation pathway, i.e. crystal growth. As we do not see evidence of crystal structure in our low-lying cluster configurations, it seems unlikely that such a large value of $\lambda$ is applicable to astrophysical dust nucleation.

%\begin{figure}[!ht]
%	\centering
%	\includegraphics[width=0.5\textwidth]{figures/ncrit_l9-5}
%	\caption{lambda 9.5}
%	\label{fig:ncrit_l9-5}
%\end{figure}
%\begin{figure}[!ht]
%	\centering
%	\includegraphics[width=0.5\textwidth]{figures/jrate_l9-0}
%	\caption{lambda 9.0}
%	\label{fig:jrate_l9-0}
%\end{figure}
%\begin{figure}[!ht]
%	\centering
%	\includegraphics[width=0.5\textwidth]{figures/ncrit_l9-0}
%	\caption{lambda 9.0}
%	\label{fig:ncrit_l9-0}
%\end{figure}
%\begin{figure}[!ht]
%	\centering
%	\includegraphics[width=0.5\textwidth]{figures/jrate_l7-0}
%	\caption{lambda 7.0}
%	\label{fig:jrate_l7-0}
%\end{figure}
%\begin{figure}[!ht]
%	\centering
%	\includegraphics[width=0.5\textwidth]{figures/ncrit_l7-0}
%	\caption{lambda 7.0}
%	\label{fig:ncrit_l7-0}
%\end{figure}

\end{document}